\documentclass[acmsmall, screen, nonacm]{acmart}

\setcopyright{acmlicensed}
\copyrightyear{2026}

\acmYear{2026}
\acmJournal{TOSEM}
\acmDOI{XXXXXXX.XXXXXXX}\renewcommand\footnotetextcopyrightpermission[1]{} 

\usepackage{nicefrac}
\usepackage{subcaption}
\usepackage{xspace}
\usepackage{graphicx}
\usepackage{amsthm}
\usepackage{multirow}
\usepackage{color}
\usepackage{pifont}
\usepackage{xcolor, colortbl}
\usepackage[breakable]{tcolorbox}
\usepackage{stfloats}

\usepackage{paralist}

\hypersetup{
colorlinks = true, 
linkcolor = black, 
citecolor = black, 
urlcolor = black 
}

\theoremstyle{definition}

\theoremstyle{definition}

\theoremstyle{definition}
\newtheorem{remark}{Remark}
\theoremstyle{definition}

\newcommand{\egovehicle}{ego-vehicle\xspace}
\newcommand{\egovehicles}{ego-vehicles\xspace}
\newcommand{\isInteraction}{\ensuremath{\mathit{isInteraction}}\xspace}
\newcommand{\AVs}{AVs\xspace} 
\newcommand{\av}{\ensuremath{\mathit{av}}\xspace} 
\newcommand{\numAVs}{\ensuremath{\mathit{m}}\xspace}
\newcommand{\timestamp}{\ensuremath{t}\xspace}
\newcommand{\firstCondInteration}{C1\xspace}
\newcommand{\secondCondInteration}{C2\xspace}
\newcommand{\cumulativeDistance}{\ensuremath{ED}\xspace}
\newcommand{\thresholdDistance}{\ensuremath{\mathit{Th_{ED}}}\xspace}
\newcommand{\fitness}{\ensuremath{\mathit{fit}}\xspace}
\newcommand{\fitnessInt}{\ensuremath{\fitness_{\mathit{int}}}\xspace}
\newcommand{\fitnessDiversity}{\ensuremath{\fitness_{\mathit{div}}}\xspace}
\newcommand{\fitnessNumCars}{\ensuremath{\fitness_{\mathit{numAVs}}}\xspace}
\newcommand{\doppelTest}{\texttt{Doppel}\-\texttt{Test}\xspace}
\newcommand{\approach}{\texttt{EVITA}\xspace}
\newcommand{\usaTwentythree}{\texttt{US101\_23}\xspace}
\newcommand{\choID}{\texttt{Cho}\xspace}
\newcommand{\usaFifteen}{\texttt{US101\_15}\xspace}
\newcommand{\cologne}{\texttt{Cologne}\xspace}
\newcommand{\usaNine}{\texttt{US101\_09}\xspace}
\newcommand{\borregas}{\texttt{Borregas Ave}\xspace}

\newcommand{\stateID}[2]{\ensuremath{\mathit{s}_{#1}^{#2}}\xspace}
\newcommand{\plannedstateID}[2]{\ensuremath{\mathit{\hat{s}}_{#1}^{#2}}\xspace}

\newcommand{\emergencySlowDownAmount}{\ensuremath{\mathit{max\_}\-\mathit{e}\-\mathit{mer}\-\mathit{gen}\-\mathit{cy\_}\-\mathit{slow}\-\mathit{down}}\xspace}
\newcommand{\minAmountAVs}{\ensuremath{\mathit{min\_}\-\mathit{AVs}}\xspace}
\newcommand{\maxAmountAVs}{\ensuremath{\mathit{max\_}\-\mathit{AVs}}\xspace}
\newcommand{\maxInitVelocity}{\ensuremath{\mathit{max\_}\-\mathit{init\_}\-\mathit{vel}}\xspace}
\newcommand{\maxTimeStep}{\ensuremath{\mathit{max\_}\-\mathit{time\_}\-\mathit{step}}\xspace}
\newcommand{\minInitVelocity}{\ensuremath{\mathit{min\_}\-\mathit{init\_}\-\mathit{vel}}\xspace}
\newcommand{\maxPositionModifier}{\ensuremath{\mathit{max\_}\-\mathit{pos\_}\-\mathit{mod}}\xspace}
\newcommand{\minPositionModifier}{\ensuremath{\mathit{min\_}\-\mathit{pos\_}\-\mathit{mod}}\xspace}
\newcommand{\minDistGoal}{\ensuremath{\mathit{min\_}\-\mathit{dist\_}\-\mathit{goal}}\xspace}
\newcommand{\minVelocity}{\ensuremath{\mathit{min\_}\-\mathit{velocity}}\xspace}
\newcommand{\maxVelocity}{\ensuremath{\mathit{max\_}\-\mathit{velocity}}\xspace}
\newcommand{\minTrafficLightDuration}{\ensuremath{\mathit{min\_}\-\mathit{green\_}\-\mathit{light\_}\-\mathit{duration}}\xspace}
\newcommand{\maxTrafficLightDuration}{\ensuremath{\mathit{max\_}\-\mathit{green\_}\-\mathit{light\_}\-\mathit{duration}}\xspace}
\newcommand{\simulationRetries}{\ensuremath{\mathit{generation\_}\-\mathit{budget}}\xspace}
\newcommand{\initialPopulation}{\ensuremath{\mathit{initial\_}\-\mathit{population\_}\mathit{size}}\xspace}
\newcommand{\distToEnd}{\ensuremath{\mathit{dist\_}\-\mathit{to\_}\-\mathit{end}}\xspace}
\newcommand{\Atwelve}{\ensuremath{\hat{A}}\textsubscript{12}\xspace}
\newcommand{\trafficLight}{\texttt{t}\xspace}
\newcommand{\planningProblem}{\texttt{p}\xspace}
\newcommand{\CommonRoad}{\texttt{Common}\-\texttt{Road}\xspace}
\newcommand{\scenario}{\individual}

\newcommand{\actualtrajectory}{\ensuremath{\mathit{trj}}\xspace}
\newcommand{\plannedtrajectory}{\ensuremath{\mathit{\hat{trj}}}\xspace}

\newcommand{\individual}{\ensuremath{\texttt{s}}\xspace}
\newcommand{\changeType}{\ensuremath{\mathit{\overline{ct}}}\xspace}
\newcommand{\frenetix}{FrenetiX\xspace}
\newcommand{\apollo}{Apollo\xspace}
\newcommand{\baidu}{Baidu Apollo\xspace}
\newcommand{\totalInts}{\texttt{Total}\-\texttt{Ints}\xspace}
\newcommand{\uniqueInts}{\texttt{UniqueInts}\xspace}
\newcommand{\totalColls}{\texttt{TotalColls}\xspace}
\newcommand{\uniqueColls}{\texttt{UniqueColls}\xspace}
\newcommand{\nVehicles}{\texttt{nVehicles}\xspace}
\newcommand{\totalFailTrafficLights}{\texttt{Total}\-\texttt{FailTrafficLights}\xspace}
\newcommand{\totalFailStop}{\texttt{Total}\-\texttt{FailStop}\xspace}
\newcommand{\totalSpeeding}{\texttt{Total}\-\texttt{Speeding}\xspace}
\newcommand{\totalScenarios}{\texttt{Total}\-\texttt{Scenarios}\xspace}
\newcommand{\validScenarios}{\texttt{Valid}\-\texttt{Scenarios}\xspace}
\newcommand{\pointOfImpact}{\texttt{POI}\xspace}
\newcommand{\relImpVel}{\texttt{RIV}\xspace}
\newcommand{\relImpAng}{\texttt{RIA}\xspace}
\newcommand{\givenNumberCars}{\ensuremath{K}\xspace}
\newcommand{\frontLeft}{\texttt{FL}\xspace}
\newcommand{\frontRight}{\texttt{FR}\xspace}
\newcommand{\frontCenter}{\texttt{FC}\xspace}
\newcommand{\backRight}{\texttt{BR}\xspace}
\newcommand{\backLeft}{\texttt{BL}\xspace}
\newcommand{\backCenter}{\texttt{BC}\xspace}
\newcommand{\middleLeft}{\texttt{ML}\xspace}
\newcommand{\middleRight}{\texttt{MR}\xspace}
\newcommand{\notApplicable}{N/A\xspace}
\newcommand{\TotalCollsBorregasEvita}{53\xspace}
\newcommand{\UniqueCollsBorregasEvita}{14\xspace}

\newcommand{\TotalCollsBorregasDopple}{30\xspace}
\newcommand{\UniqueCollsBorregasDopple}{4\xspace}

\newcounter{resq}[subsection]
\newenvironment{resq}[1][]{
\refstepcounter{resq}
\par\medskip
\noindent \textbf{RQ\theresq.}~\textit{#1} \rmfamily \\}{\smallskip}

\newcommand{\prevquestion}{RQ\the\numexpr\value{resq}-1\relax\xspace}
\usepackage{todonotes}

\begin{document}

\title{Automatic Testing of Interacting Autonomous Vehicles}

\author{Fabio Cavaleri}
\affiliation{
\institution{Universit{\`a} della Svizzera Italiana}
\city{Lugano}
\country{Switzerland}
}
\orcid{0009-0008-2737-5212}
\email{fabio.cavaleri@usi.ch}
\authornote{Corresponding author}

\author{Alessio Gambi}
\affiliation{
\institution{AIT - Austrian Institute of Technology}
\city{Vienna}
\country{Austria}
}
\affiliation{
\institution{The Italian Institute of Artificial Intelligence (AI4I) }
\city{Turin}
\country{Italy}
}
\orcid{0000-0002-0132-6497}
\email{alessio.gambi@ait.ac.at}

\author{Paolo Arcaini}
\affiliation{
\institution{National Institute of Informatics}
\city{Tokyo}
\country{Japan}
}
\orcid{0000-0002-6253-4062}
\email{arcaini@nii.ac.jp}

\author{Dejan Ni\v{c}kovi{\'c}}
\affiliation{
\institution{AIT - Austrian Institute of Technology}
\city{Vienna}
\country{Austria}
}
\orcid{0000-0001-5468-0396}
\email{dejan.nickovic@ait.ac.at}

\author{Mauro Pezz{\`e}}
\affiliation{
\institution{USI Università della Svizzera Italiana}
\city{Lugano}
\country{Switzerland}
}
\affiliation{
\institution{Universit{\`a} degli Studi di Milano-Bicocca}
\city{Milano}
\country{Italy}
}
\orcid{0000-0001-5193-7379}
\email{mauro.pezze@usi.ch}

\renewcommand{\shortauthors}{Cavaleri et al.}

\keywords{interaction testing, feature interaction, search-based testing, scenario-based testing}

\begin{abstract}
Autonomous vehicles (\AVs) must be thoroughly tested to meet high safety standards and avoid endangering both AV passengers and road users. 
Scenario-based testing implements driving scenarios in virtual simulation environments as a cost-effective alternative to field testing. 
Common scenario-based testing approaches set the environment and the surrounding traffic and test a single AV. 
Recent studies show that the approaches that test single \AVs miss critical behaviors that emerge from interactions among multiple \AVs. 
Effective approaches to test scenarios that emerge from n-way interactions must address the combinatorial explosion that the presence of multiple \AVs further exacerbates.

In this paper, we propose \approach, an approach that leverages multi-objective optimization to generate scenarios that trigger multiple and diverse \AVs interactions, while minimizing the complexity of the generated scenarios, to effectively test multiple interacting \AVs and reveal safety-critical scenarios that current approaches overlook.
The experimental results that we discuss in this paper confirm that \approach triggers a higher variety of \AVs interactions than state-of-the-art approaches, thus improving the likelihood to reveal safety-critical behaviors. 
%
\end{abstract}

\maketitle

\section{Introduction}
\label{sec:introduction}
Autonomous vehicles (\AVs) are increasingly tested in the real world~\cite{Bansal2017}, both in \emph{driverless} mode or with a safety driver on board~\cite{dmv}.
At the time of writing, there are more than $1,400$ \AVs of over $80$ manufacturers that test the vehicles across the USA~\cite{techcrunch}, and $16,000$ test licenses for autonomous vehicles issued in China~\cite{gov-cn}.
The physical testing of \AVs in the real world is constrained by regulatory and environmental factors, and is expensive~\cite{kalra2016driving}.
\AVs must abide by road signs (static elements) and traffic lights (dynamic elements), and must coordinate with other vehicles (reactive elements) driven by humans and other \AVs, to avoid dangerous interactions~\cite{china-tests}, hazardous situations~\cite{guardian,cnnCruise}, and large-scale disruptions, such as roadblocks~\cite{thevergeDriverlessCruise,bbc-waymo}. 

Testing of \AVs in simulation has been widely adopted in academia and industry as a cost-effective alternative to field testing~\cite{Schoner2018,DBLP:conf/sigsoft/LouDZZ022} and as the only viable solution for testing \AVs under conditions unlikely to be observed or too dangerous to reproduce in the real world.

\begin{figure}[t]
\centering
\begin{subfigure}{0.8\textwidth}
\centering
\includegraphics[width = \textwidth]{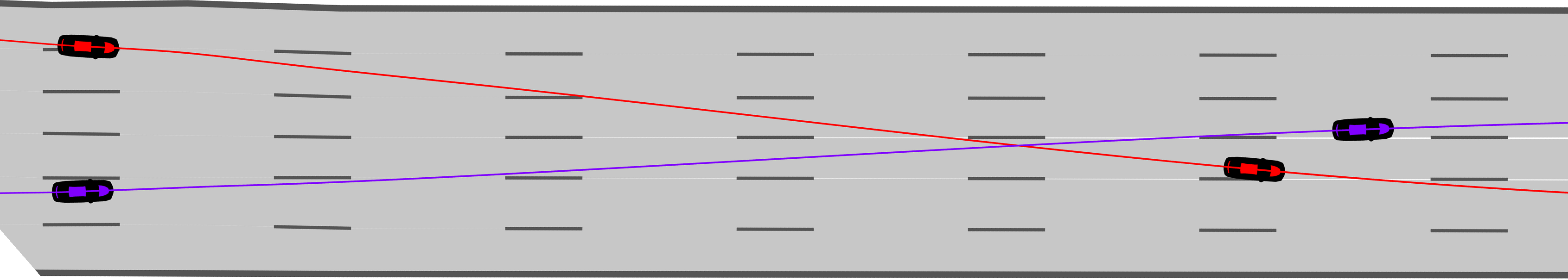}
\caption{The \av (in red) avoids the collision with the NPC (in purple)}
\label{fig:motivating-example-a}
\end{subfigure}
\hfill
\begin{subfigure}{0.8\textwidth}
\centering
\includegraphics[width = \textwidth]{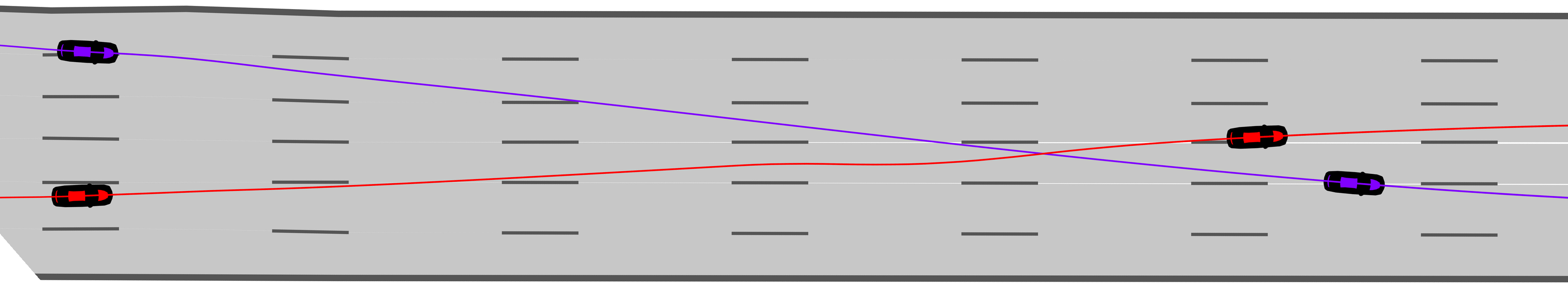}
\caption{The \av (in red) avoids the collision with the NPC (in purple)}
\label{fig:motivating-example-b}
\end{subfigure}
\hfill
\begin{subfigure}{0.8\textwidth}
\centering
\includegraphics[width = \textwidth]{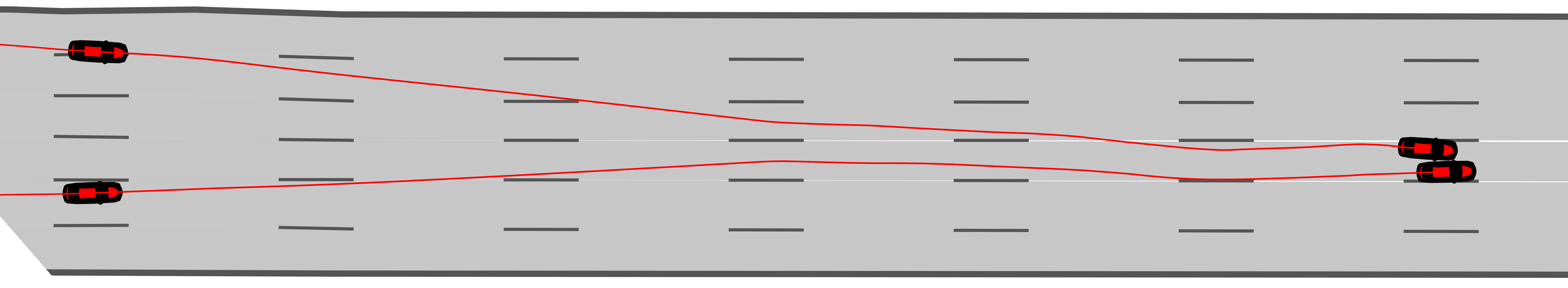}
\caption{The two \AVs (in red) collide}
\label{fig:motivating-example-c}
\end{subfigure}
\caption{Motivating example: Testing two vehicles with the same initial and final positions, and trajectories}
\label{fig:motivating-example}
\end{figure}

Scenario-based testing creates driving scenarios that challenge the \AVs under test, referred to as the \emph{\egovehicles}, to expose misbehavior that results in collisions and traffic law violations~\cite{Koopman2016}.
State-of-the-art scenario-based testing approaches target single \AVs. 
They test an \av against pre-programmed Non-Playable Characters (NPCs) that are road users, either completely non-interactive or behaving according to simplistic models~(\cite{IDM,MIDM,TianASE2022}). 
They rarely involve other scenario elements, such as traffic lights and road signs~\cite{AVTesting-Survey}.
Testing a single \av against pre-programmed NPCs does not assess the mutual reactions of multiple vehicles on the road~\cite{albeaik2022limitations} and, consequently, overlooks issues that may cause accidents that state-of-the-art approaches miss~\cite{doppleganger}.
Figure~\ref{fig:motivating-example} exemplifies the critical scenarios that emerge from the mutual interactions among \AVs. 
Figures~\ref{fig:motivating-example-a} and ~\ref{fig:motivating-example-b} show that the \av (in \textcolor{red}{red} in the figures) avoids the collision with the NPC (in \textcolor{violet}{purple} in the figures), while two \AVs starting with the same initial and final positions do not avoid the collision, due to their mutual interactions (Figure~\ref{fig:motivating-example-c}).
The example shows that the behavior of the \egovehicle depends on its expectations about surrounding traffic behavior, and single-AV scenarios fail to capture when \AVs adjust their trajectories in response to nearby vehicles.
As the number and diversity of \AVs on public roads steadily increase~\cite{Bansal2017, WEF-2025-AVs-report}, the chances of mutually interacting \AVs that lead to hazardous situations highly increase. 

In this paper, we address the shortcomings of single-AV scenario-based testing by introducing \emph{interaction testing}, a methodology to generate critical scenarios that systematically test interactions influencing the behavior of the \AVs under test.
Testing \AVs against system-level interactions can reveal hazards that arise from emergent behaviors triggered by a difficult-to-predict interplay among multiple influencing factors, and that would otherwise go undetected. 
Multi-\AVs scenarios provide a more realistic approximation of traffic behavior than single-\AVs scenarios, in which simple, inflexible logic controls the road users.

We introduce \approach, \emph{EVolutionary Interaction Testing Approach}, an interaction testing approach for automatically generating critical scenarios involving multiple \AVs in both highway and urban traffic, with traffic lights and road signs.
\approach leverages multi-objective optimization to generate scenarios that maximize the number and variety of interactions, while reducing scenario complexity by keeping the number of \AVs to a minimum.
%
\approach aims to expose safety-critical \mbox{(mis-)interactions} and collisions that involve multiple interacting \AVs and that might be difficult to replicate using pre-programmed behaviors in single-AV scenarios.

We evaluate \approach against two experimental setups: \frenetix for highways and suburban roads, and \baidu for urban scenarios.
\frenetix~\cite{Frenetix} is an open-source, highly performing motion planner that we tested in scenarios using replicas of real-life highways and suburban roads.
We experimented with the \frenetix multi-agent platform~\cite{multiagent2024} developed for \CommonRoad~\cite{Althoff2017}, an established and well-maintained simulation framework widely used in research.\footnote{See the \CommonRoad Web Page at \url{https://commonroad.in.tum.de}}
\baidu~\cite{apollo} is an industrial-scale autonomous driving agent that we use to evaluate \approach in complex urban scenarios featuring intersections, traffic lights, and road signs.
We extend Huai~et~al.'s \baidu framework~\cite{doppleganger} adopted by \doppelTest, with additional interaction monitors, and new data analyses.

The results that we discuss in Section~\ref{sec:evaluation} show that \approach consistently triggers diverse \AVs interactions and identifies a broad range of critical scenarios, outperforming \doppelTest, the only state-of-the-art open-source approach that, to the best of our knowledge, can generate scenarios involving multiple \AVs.

This paper contributes to testing \AVs by
\begin{itemize}
\item defining interaction testing, a methodology for exploring emergent behaviors resulting from the interplay of \AVs and other traffic-influencing factors;
\item proposing \approach, an interaction testing approach to systematically generate scenarios that involve \AVs interactions in driving simulations;
\item presenting an extensive evaluation involving multiple simulators, \AVs systems, and diverse road environments.
\end{itemize}
\approach's source code, data, and scripts to replicate the experiments are available in the replication package~\cite{replication-package-evita}.

The remainder of this paper is organized as follows:
Section~\ref{sec:background} introduces background concepts to make the paper self-contained.
Section~\ref{sec:interactions} introduces our novel notion of interactions, central to \approach, which is presented in Section~\ref{sec:methodology}.
Section~\ref{sec:evaluation} presents the evaluation methodology, including research questions, experimental settings, and executed experiments.
Section~\ref{sec:results} presents the obtained results.
Section~\ref{sec:threatsToValidity} discusses threats to the validity of our conclusions.
Section~\ref{sec:related} reviews related work, before Section~\ref{sec:conclusions} concludes the paper.

\section{Multi-AV CommonRoad Driving Scenarios}
\label{sec:background}
The \emph{COMposable benchmarks for MOtioN planning on ROADs} (\CommonRoad) framework is an established, open-source framework that supports researchers in developing and validating motion planners, the core components of autonomous vehicles~\cite{Althoff2017}.
It offers open access to a large database of synthetic and realistic traffic scenarios~\cite{commonRoadScenarios} and the possibility to import scenarios defined in other standards, such ASAM OpenScenario~\cite{OpenSCENARIO-CommonRoad} and OpenDrive~\cite{OpenDRIVE-CommonRoad}.

\begin{figure}
\centering
\includegraphics[width = \linewidth]{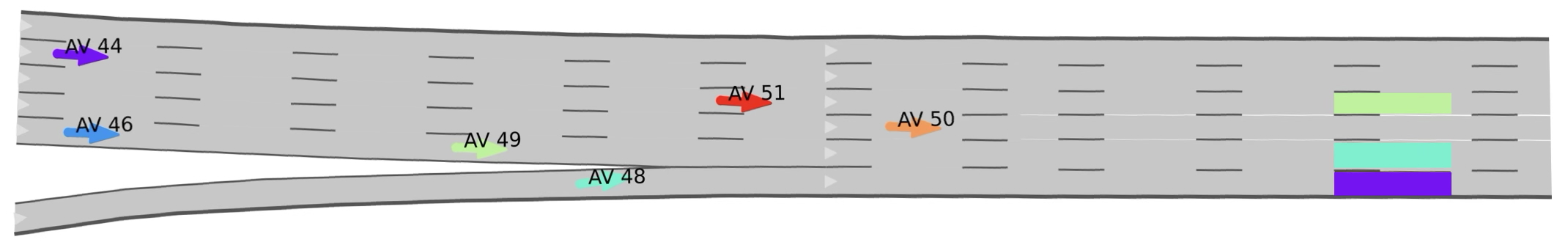}
\caption{Example of a Multi-AV CommonRoad highway scenario}
\label{fig:multi-av-scenario}
\end{figure}

Figure~\ref{fig:multi-av-scenario} shows an example of a Multi-AV CommonRoad highway scenario. 
The colored arrows represent the initial position and direction of the \AVs, whereas the colored rectangles identify the goal areas each \av must reach. 
Some goal areas overlap; thus, not all goal areas are visible in the figure.
\CommonRoad's multi-AV scenarios comprise 
\begin{inparaenum}[(i)]
\item a \textbf{road network} encoded in the Lanelet2~\cite{lanelets2} format that describes the ``drivable'' areas of the map as lanelets that are polygons linked together to form the network (tiny light gray triangles indicate the direction of the traffic in the figure), 
\item a set of \textbf{obstacles} representing static objects and dynamic traffic participants having geometric and kinematic properties (for instance \AVs during simulation),
\item a set of scenario elements such as \textbf{line markings}, \textbf{road signs}, and \textbf{traffic lights} (not present in the figure),
\item a set of \textbf{planning problems} that encode driving tasks.
Each planning problem is assigned to an \av and specifies both the initial state \stateID{av}{\timestamp_{0}} and the target states. 
States have attributes defining each \av's initial or expected position, rotation, velocity, and acceleration, which can be either precise values or intervals.
In Figure~\ref{fig:multi-av-scenario}, the initial states are represented as colored arrows with precise position and direction, whereas the target states are depicted as rectangles that identify intervals over longitudinal and lateral positions, with matching colors.
\end{inparaenum}

\approach generates multi-AV scenarios by specifying for each \av:
\begin{inparaenum}[(i)]
\item the initial position, rotation, and speed as values; and
\item the target position and timeout as intervals. 
\end{inparaenum}    
\approach defines a conservative timeout to complete the driving tasks (\maxTimeStep), to avoid over-constraining the planning problems. 
It ensures that the \AVs start on the road and that their planning problems are satisfiable, to avoid generating invalid or irrelevant scenarios (see Section~\ref{sec:constraints}).

\CommonRoad works with logical time and unfolds the scenario simulations in discrete \emph{simulation steps}. 
\approach uses the default simulation step duration of $0.1$ seconds.
During simulation, \approach monitors the current state of every \av (\stateID{av}{\timestamp_{i}}), which includes the physical properties of the \av, position, rotation, and speed at each simulation step $\timestamp_{i}$. 
By the end of the simulation, the trajectory that each \av followed is the sequence of its observed states,  
$\actualtrajectory_{av} = [\stateID{}{\timestamp_{0}}, \ldots, \stateID{}{\timestamp_{N}}]_{av}$.
\approach also observes the trajectories each \av has planned at different times.
At time $\timestamp_{i}$, the sequence of states planned by the \av over the planning horizon $H$ is 
$\plannedtrajectory^{\timestamp_{i}}_{av} = [\plannedstateID{\timestamp_{i}}{\timestamp_{i+1}}, \ldots, \plannedstateID{\timestamp_{i}}{\timestamp_{i+H}}]_{av}$.
\approach identifies \AVs interactions by comparing the trajectories the \AVs planned during the scenario.

\section{Identifying AVs' Interactions} 
\label{sec:interactions}
Autonomous vehicles implement a sense-plan-act (SPA) closed-loop architecture to drive while continuously interacting with other vehicles and reacting to scenario elements.
According to this architecture,
\begin{inparaenum}[(1)]
\item \AVs continuously perceive their surroundings using sensors, like cameras, LiDARs, and radars, and fuse raw sensor data with map information to localize themselves and identify relevant objects around them;
\item based on localization and object detection data, \AVs identify a path to reach a target location on the map and plan a trajectory to drive there safely; finally, 
\item low-level controllers translate the planned trajectories into driving commands (e.g., throttle, steering) that control the vehicle's movement.
\end{inparaenum}

During the SPA loop, \AVs can \emph{follow} the planned trajectories, \emph{adjust} them to improve performance or safety, or \emph{replan} them entirely in response to endogenous factors. 
For instance,
\begin{inparaenum}[(i)]
\item the \av might simply follow its trajectory, while driving straight on an empty road,
\item it might adjust its trajectory to improve passenger comfort, while taking a turn,
\item it might promptly plan to stop, before reaching the traffic light, after sensing a red light.
\end{inparaenum}

\approach privileges changes of trajectories, to systematically generate test scenarios that yield diverse interactions, and challenge the \AVs in both foreseeable and unforeseen situations that may lead to dangerous outcomes and violations of traffic rules, as in the example of Figure~\ref{fig:motivating-example}.
\approach identifies interactions among the \AVs and between the \AVs and other scenario elements, and quantitatively analyzes how the interactions affect the \AVs' behavior by comparing consecutive planned trajectories (\plannedtrajectory). 
\approach identifies an interaction of an \av at time $\timestamp_{i}$ if:

\smallskip
\noindent\textbf{Condition \firstCondInteration}\label{cond:one}: \av replanned its trajectory between $\timestamp_{i-1}$ and $\timestamp_{i}$, and

\noindent\textbf{Condition \secondCondInteration}\label{cond:two}: the replanning is likely caused by a scenario element.
\smallskip

The algorithm for checking condition \firstCondInteration depends on the way \apollo and \frenetix recompute the trajectories; the algorithm for checking condition \secondCondInteration depends on the obstacles in the scenarios (only cars in \frenetix; cars, traffic lights, and road signs in \apollo).

\subsection{Identifying Trajectory Replanning (Condition C1)}
\apollo adopts a modular architecture, which implements sensing, planning, and control as autonomous components~\cite{DBLP:journals/access/CarvalhoSYFK25} that communicate via the message-passing middleware ROS2~\cite{DBLP:journals/scirobotics/MacenskiFGLW22}.
The \textit{sensing} component of \apollo sends messages with data about the perceived objects to the \textit{planning} component. 
The \textit{planning} component uses the data to compute new trajectories, and it sends them to the \textit{control} module.
\approach identifies a trajectory replanning in \apollo from the messages of the planning module.

\frenetix recomputes a new trajectory at every control cycle, even when no different trajectory is necessary.
\approach identifies a trajectory replanning by comparing the trajectory $\plannedtrajectory^{\timestamp_{i}}_{av}$ computed at time $\timestamp_{i}$  against the trajectory $\plannedtrajectory^{\timestamp_{i-1}}_{av}$ computed at time $\timestamp_{i-1}$ for the same vehicle \av as follows:
\begin{enumerate}
\item it trims the sub-trajectories with same timestamps in $\plannedtrajectory^{\timestamp_{i-1}}_{av}$ and $\plannedtrajectory^{\timestamp_{i}}_{av}$:
\begin{center}
$[\plannedstateID{\timestamp_{i-1}}{\timestamp_{i+1}},$
$\ldots,$
$\plannedstateID{\timestamp_{i-1}}{\timestamp_{i+H}}]_{av}$
and 
$[\plannedstateID{\timestamp_{i}}{\timestamp_{i+1}},$
$\ldots,$
$\plannedstateID{\timestamp_{i}}{\timestamp_{i+H}}]_{av}$,
\end{center}
\item it computes the cumulative Euclidean distance between pairs of \av's predicted positions following Magdy et al.~\cite{Magdy2015}:
\begin{equation}
\cumulativeDistance = \sum_{j=i+1}^{i+H} \left \lVert \plannedstateID{\timestamp_{i-1}}{\timestamp_{j}}\text{.position} - \plannedstateID{\timestamp_{i}}{\timestamp_{j}}\text{.position}] \right \rVert_2
\end{equation}
\item it flags a significant difference between the predicted trajectories when $\cumulativeDistance$ exceeds the threshold \thresholdDistance.
\end{enumerate}

\begin{figure}[t]
\centering
\includegraphics[width = \textwidth]{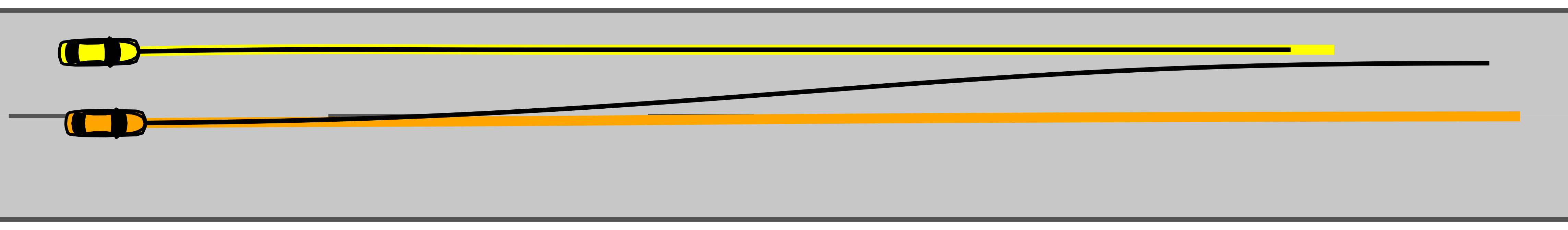}
\caption{Identifying trajectory replanning in \frenetix}
\label{fig:comparisonPrevAndNewCompTrajs}
\end{figure}
Figure~\ref{fig:comparisonPrevAndNewCompTrajs} illustrates how \approach identifies a trajectory replanning for \frenetix.
The figure shows the trajectory of the \AVs at time $\timestamp_{i-1}$ in black, respectively, and at time $\timestamp_{i}$ in yellow and orange. 
\approach does not identify a trajectory replanning for the yellow vehicle on the left (upper) lane, since the cumulative Euclidean difference between the trajectories planned at $\timestamp_{i-1}$ and $\timestamp_{i}$ does not exceed the threshold \thresholdDistance. 
It identifies a trajectory replanning for the orange vehicle on the right (bottom) lane, since the trajectories planned at $\timestamp_{i-1}$ and $\timestamp_{i}$ differ significantly and their cumulative Euclidean difference surpasses the threshold \thresholdDistance . 

\subsection{Identifying Scenario Elements Causing Replanning (Condition \secondCondInteration)}
A series of consecutive trajectory replannings (\firstCondInteration) may indicate either a reaction of an \av to some scenario elements or a routine replanning. 
Reactions to some scenario elements include situations when the \egovehicle is approaching another vehicle like in Figure~\ref{fig:comparisonPrevAndNewCompTrajs}, a traffic light or a road sign.
Routine replannings occur in normal conditions, for example, \apollo generates replanning messages while taking turns to adjust its trajectory, whereas \frenetix replans if its reference path requires a lane change. 

\approach prunes ``spurious'' trajectory changes, following Tien~et~al.'s suggestions~\cite{TianASE2022} of seeking a likely explanation for the observed significant trajectory change.
\approach checks for the presence of obstacles, cars, traffic lights, road signs, in correspondence of a relevant trajectory change, like in the case of Figure~\ref{fig:comparisonPrevAndNewCompTrajs}.

For \baidu, \approach relies on both \apollo's planning and perception module.
The planning module associates a decision message with the replanned trajectory to indicate the reason for replanning and the responsible scenario element or vehicle. 
The perception module generates messages when it identifies obstacles (for instance, other vehicles, traffic lights, including their color, and road signs, like stop and yield signs).
For \frenetix, \approach relies on the underlying multi-AV \CommonRoad execution platform~\cite{multiagent2024}, which provides (ground truth) information about vehicles, traffic lights, and road signs during the execution of the scenarios.

\subsection{Classifying Interactions}
\label{sec:Classifying-Interactions}
\begin{figure}[!t]
\centering
\begin{subfigure}{0.49\textwidth}
\centering
\includegraphics[width=0.4\linewidth]{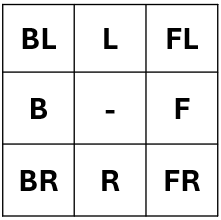}
\caption{Possible types of change of trajectory}
\label{fig:interaction-map-classes}
\end{subfigure}
\hfill
\begin{subfigure}{0.49\textwidth}
\centering
\includegraphics[width=0.5\linewidth]{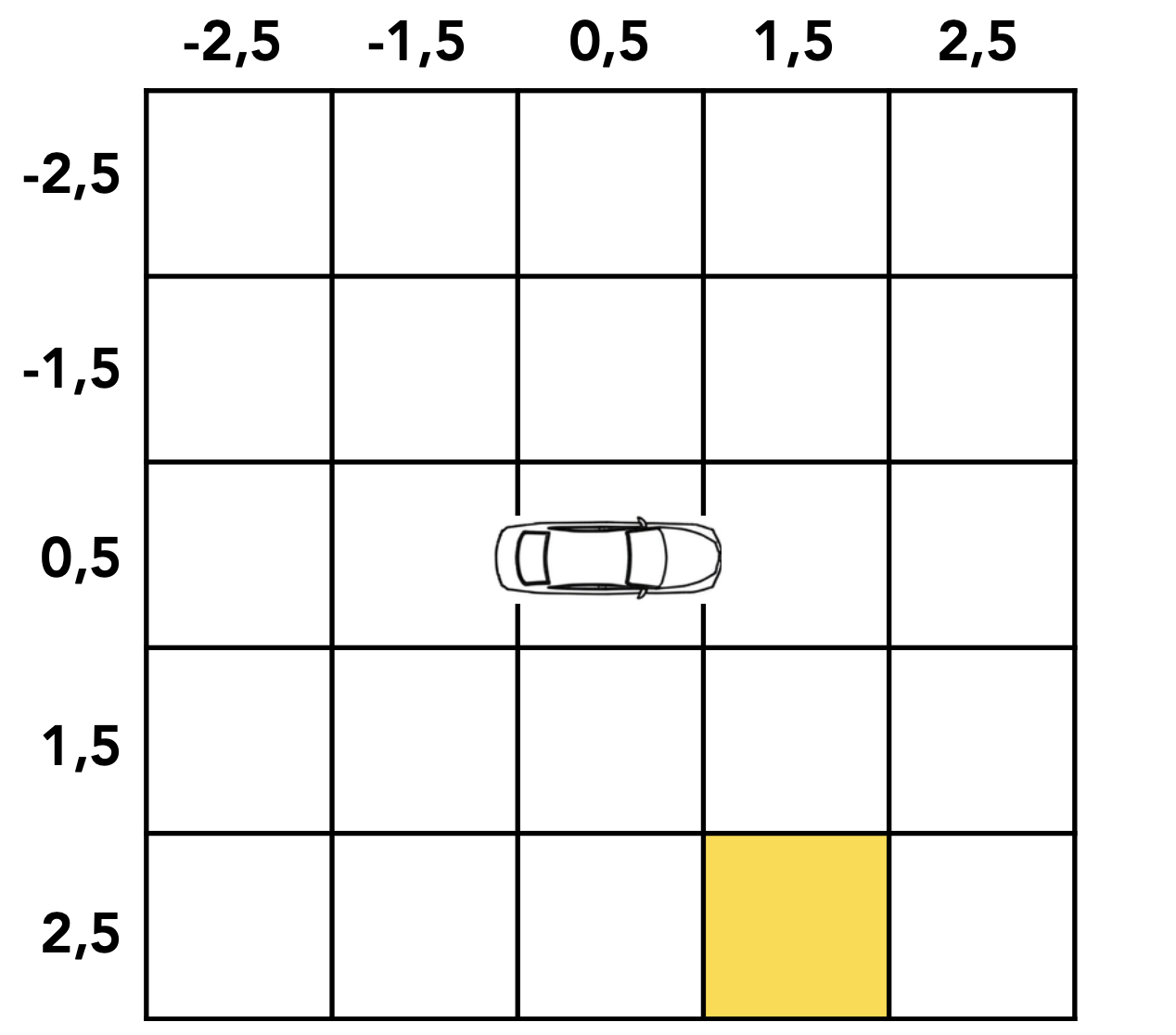}
\caption{Interaction map}
\label{fig:interaction-map-example}
\end{subfigure}
\caption{The interaction maps from the trajectory replanning of the orange vehicle in Figure~\ref{fig:comparisonPrevAndNewCompTrajs}}
\label{fig:interaction-maps}
\end{figure}

\approach classifies identified interactions as {\it forward (F)}, {\it forward-left (FL)}, {\it forward-right (FR)}, {\it left (L)}, {\it right (R)}, {\it backwards (B)}, {\it backwards-left (BL)}, and {\it backwards-right (BR)}, depending on the relative change in the planned trajectory.
It uses \emph{interaction maps} (see~Figure~\ref{fig:interaction-maps}) that adapt the feature maps proposed by Riccio et al.~\cite{ZohdinasabISSTA2021}, for assessing how many types of interaction a scenario covers.

\approach builds the interaction map for an $\av$ by
\begin{enumerate}
\item  extracting the last planned states having the same timestep ($[\plannedstateID{\timestamp_{i-1}}{\timestamp_{i+H}}]_{av}$ and $[\plannedstateID{\timestamp_{i}}{\timestamp_{i+H}}]_{av}$) from the planned trajectories defining the interaction ($\plannedtrajectory^{\timestamp_{i-1}}_{av}$, $\plannedtrajectory^{\timestamp_{i}}_{av}$),
\item computing the difference between the position of these states in the longitudinal and lateral directions, thereby quantifying how much the original trajectory changed due to the interaction,
\item classifying the difference as {\it forward (F)}, {\it forward-left (FL)}, {\it forward-right (FR)}, {\it left (L)}, {\it right (R)}, {\it backwards (B)}, {\it backwards-left (BL)}, and {\it backwards-right (BR)}, depending on the relative change in the planned trajectory,
\item building a two-dimensional grid whose cells capture both the type and the degree of changes resulting from the interaction. 
\end{enumerate}

Figure~\ref{fig:interaction-map-classes} illustrates the nine possible changes of trajectory. 
Figure~\ref{fig:interaction-map-example} shows the interaction map resulting from the trajectory replanning of the orange vehicle in Figure~\ref{fig:comparisonPrevAndNewCompTrajs}. 
The interaction map is a two-dimensional grid, where the central cell represents the ``no-trajectory-change'', the cells farther from the central cell identify interactions with larger effects, and the colors represent the number of times each interaction type has been observed. 
Light colors correspond to none (white) or few observations (yellow), whereas darker colors (black) correspond to many observations. 
In the figure, we observe few interactions (yellow) with the Right/FrontRight (R/FR) cell, which indicate the interactions between the two vehicles of Figure~\ref{fig:comparisonPrevAndNewCompTrajs}, caused by the change of trajectory of the orange vehicle that ``challenges'' the yellow vehicle on the right, front-right.

By mapping each interaction to a specific cell in the interaction map, \approach measures the number of interaction types a scenario exercises and assesses its overall effect on \AVs' behavior.
\approach determines the sensitivity of the analysis by adjusting the grid size: a larger number of cells makes the analysis more sensitive to small changes; a smaller number of cells reduces the map's ability to distinguish scenarios that cause different interactions.

We define the interval of lateral displacement as twice the width of a standard lane (ca. $3.70$m) to account for lane changes and define the interval of longitudinal displacement as twice the length of a standard vehicle (ca. $4.5$m), based on the observation that changes in \AVs' trajectories are limited by the vehicles' physics and common sense.
We split longitudinal and lateral dimensions into a fixed number of bins ($10$) and add an extra bucket to account for edge cases. 

\section{Testing \AVs' Interactions with \approach}
\label{sec:methodology}
\approach generates multi-AV scenarios that expose safety-critical issues~\cite{BehAVExplor} in \CommonRoad format.
\approach's scenarios test multiple \AVs against behaviors that emerge from the interplay between the \AVs under test and other scenario elements, including dynamic traffic lights and static road signs.
\approach improves the effectiveness of testing by generating different scenarios that systematically explore the space of possible interactions affecting the \av behavior.
This section describes how \approach generates diverse scenarios and maximizes both the number and types of interactions by reformulating scenario generation as a multi-objective optimization problem.

\subsection{Search Algorithm}
\label{sec:searchAlgorithm}
Following Tian et al.'s observation~\cite{TianASE2022} that many critical scenarios are characterized by different interactions among \AVs and between \AVs and other scenario elements, \approach generates scenarios with many different interactions using a search algorithm.

\approach generates multi-AV scenarios by configuring the timing of traffic light cycles and the autonomous vehicles for a given map. \approach relies on \doppelTest's traffic lights control that ensures the correct synchronization of traffic signals at each intersection.
\approach configures autonomous vehicles by maximizing the number and the variety of interactions while minimizing the number of \AVs per scenario, thereby focusing on the \AVs relevant to testing, by adopting a multi-objective search algorithm. 
\approach uses NSGA-II (Non-Dominated Sorting Genetic Algorithm II)~\cite{deb2002fast}, a state-of-the-art multi-objective algorithm that finds many applications in different domains~\cite{doppleganger,Abdessalem2018TVC,stretchIV2023,avoidCollICST2020,MOSAT,driveCharICST2021,variationsADSsTOSEM2025,HumeniukSCICO2023,ZohdinasabRT24}. 

In a nutshell, NSGA-II evolves a population of individuals (Section~\ref{sec:individuals}) that represent multi-AV scenarios.
After creating the initial population of individuals (Section~\ref{sec:search-initialization}), NSGA-II iteratively
\begin{inparaenum}[(1)]
\item calculates the fitness functions (Section~\ref{sec:fitnessFunctions}) for each individual, 
\item ranks the individuals and calculates the crowding distance,
\item evolves the population by generating new individuals via crossover and mutation operators (Section~\ref{sec:evolution}).
\end{inparaenum}
\approach iterates the evolution loop up to exhausting a search budget. 

\subsection{Individuals}
\label{sec:individuals}
As discussed in Section~\ref{sec:background}, a multi-AV \CommonRoad scenario consists of a road network, which may include traffic lights and road signs, and a set of planning problems that define the driving tasks for the autonomous vehicles. 
\approach maximizes the interactions by keeping the map fixed while both mutating and recombining the set of planning problems and the timing of traffic signals. 
\approach encodes each \emph{individual} scenario as a sequence of attributes representing a variable number of planning problems ranging from $2$ to $\numAVs$ and a fixed number of traffic lights (between $0$ and $l$, as defined by the map):  $\scenario = \{\planningProblem_{\av_1}, \ldots, \planningProblem_{\av_{\numAVs}}; \trafficLight_{j}, \ldots, \trafficLight_{l}\}$

Each planning problem $\planningProblem_{\av_i}$ has attributes that define the initial position and velocity of the \av, as well as the position of the goal area. 
\approach places vehicles on the roads and computes their initial rotation based on the road curvature at the initial position, ensuring the vehicles always follow the direction of the road ahead. 
The initial velocity is a non-negative real number, which corresponds to the vehicle's initial speed in meters per second. 
\approach places the goal areas at the end of the roads to maximize travel distance. \approach relies on the \CommonRoad representation of goal areas as rectangles centered in the given position and spanning the width of the underlying lane. \approach executes a planning problem by assigning it to an \av under test, which shall drive from the initial position to the goal area without colliding with other vehicles, driving off-road, or violating any other traffic rule, for instance, speeding, running at a red light, or failing to stop at a stop sign.

Each traffic light $\trafficLight_{j}$ has attributes that define its initial color and the intervals for transitioning from one color to the next one. 
During the execution, the underlying simulation platform ensures that colors transition as prescribed.

\subsection{Search Initialization}
\label{sec:search-initialization}
\approach initializes the search by generating an initial population of \initialPopulation random individuals. 
Each individual contains a random number of planning problems ranging from \mbox{\minAmountAVs} to \maxAmountAVs, initialized by uniformly sampling the initial and target positions and the speed values. \approach randomly initializes the colors and the green light intervals of the traffic lights in the map. 
The duration of the green light ranges from \mbox{\minTrafficLightDuration} to \mbox{\maxTrafficLightDuration} seconds, whereas the duration of the yellow light has a fixed duration of $3.0$ seconds. 
\approach configures the traffic lights to wait an additional $2.0$ seconds before turning to green to enable clearing the intersection.
To simplify the problem without loss of generality, \approach forces all traffic lights to have the same green light duration.

Table~\ref{tab:confValues} at page~\pageref{tab:confValues} lists all the configurable \emph{hyperparameters} of the algorithm, including \mbox{\minAmountAVs}, \maxAmountAVs, \minTrafficLightDuration, \maxTrafficLightDuration, and \mbox{\initialPopulation}.

\subsection{Evolution}
\label{sec:evolution}
The search algorithm evolves the scenarios by incrementally mutating and recombining the planning problems $\planningProblem_{\av}$ 
and the traffic lights configuration using mutation and crossover operations.

\subsubsection{Mutation}
\approach mutates the attributes of each planning problem, the set of planning problems, and the initial color and green light duration of traffic lights.
\approach mutates the planning problem $\planningProblem_{\av}$ associated to each \av in \AVs with a probability of $1/|\AVs|$ and its attributes with a probability of $1/3$ as follows:
\begin{itemize}
\item \textbf{Initial Position}: \approach mutates the initial position of \av by either adding a random value in the range $[\minPositionModifier, \maxPositionModifier]$\footnote{\minPositionModifier and \maxPositionModifier are search hyperparameters (see Table~\ref{tab:confValues}).} or selecting a different initial lane.
\item \textbf{Initial Velocity}: \approach mutates the initial velocity of \av by a random value in the range $[\minInitVelocity,\maxInitVelocity]$\footnote{\minInitVelocity and \maxInitVelocity are search hyperparameters (see Table~\ref{tab:confValues}).}.
\item \textbf{Goal Area}: \approach mutates the goal area of \av by moving the goal area to a different lane.
\end{itemize}
\approach adds or removes planning problems, thus autonomous vehicles, from a multi-AV scenario with a probability that depends on the number of vehicles in it:
\begin{itemize}
\item \textbf{Add AV}: \approach adds a newly generated planning problem $\planningProblem_{\av^\prime}$ to the scenario, unless the scenario contains \maxAmountAVs autonomous vehicles.
\item \textbf{Remove AV}: \approach removes a planning problem randomly, unless the scenario contains \minAmountAVs autonomous vehicles.
\end{itemize}
\approach mutates traffic lights attributes with a probability of $1/2$ as follows:
\begin{itemize}
\item \textbf{Green light duration}: \approach assigns a new random duration (between \minTrafficLightDuration and \maxTrafficLightDuration) to all traffic lights' green duration.
\item \textbf{Initial color}: \approach changes the initial color of all traffic lights, ensuring that colors are consistent with traffic regulations.
\end{itemize}

\subsubsection{Crossover}
\approach recombines the information about planning problems between two individuals using a customized crossover operator. 
Given two scenarios:
\begin{equation}\label{eq:parents}
\{\planningProblem^1_{\av_1}, \ldots, \planningProblem^1_{\av_i}, \planningProblem^1_{\av_{i+1}}, \ldots, \planningProblem^1_{\av_{r}}
; \trafficLight^1_{j}, \ldots, \trafficLight^1_{m}\}
\qquad
\{\planningProblem^2_{\av_1}, \ldots, \planningProblem^2_{\av_k}, \planningProblem^2_{\av_{k+1}}, \ldots, \planningProblem^2_{\av_{s}}
; \trafficLight^2_{j}, \ldots, \trafficLight^2_{m}\}
\end{equation}
\approach splits their planning problem sets at random positions ($i$ and $k$) and recombines them to form two new individuals:
\[
\{\planningProblem_{\av^1_1}, \ldots, \planningProblem_{\av^1_i}, \planningProblem_{\av^2_{k+1}}, \ldots, \planningProblem_{\av^2_{s}}
; \trafficLight^1_{j}, \ldots, \trafficLight^1_{m}\}
\qquad
\{\planningProblem_{\av^2_1}, \ldots, \planningProblem_{\av^2_k}, \planningProblem_{\av^1_{i+1}}, \ldots, \planningProblem_{\av^1_{r}}
; \trafficLight^2_{j}, \ldots, \trafficLight^2_{m}\}
\]

Alternatively, \approach recombines the information about traffic lights between two parent scenarios by swapping their configurations. 
Thus, given two scenarios as in Eq.~\ref{eq:parents}, it produces two new individuals:
\[
\{\planningProblem^1_{\av_1}, \ldots, \planningProblem^1_{\av_i}, \planningProblem^1_{\av_{i+1}}, \ldots, \planningProblem^1_{\av_{r}}
; \trafficLight^2_{j}, \ldots, \trafficLight^2_{m}\}
\qquad
\{\planningProblem^2_{\av_1}, \ldots, \planningProblem^2_{\av_k}, \planningProblem^2_{\av_{k+1}}, \ldots, \planningProblem^2_{\av_{s}}
; \trafficLight^1_{j}, \ldots, \trafficLight^1_{m}\}
\]

\subsection{Scenario Validation}
\label{sec:constraints}
\approach relies on random individual generation and evolution, which may generate \emph{invalid} and \emph{non-interesting} scenarios.
Invalid scenarios are situations that cannot be solved. For instance, an \av with a goal area that is not reachable from the initial state of the \av, or two autonomous vehicles spawned in the same initial position. 
Not interesting scenarios are situations irrelevant to testing. For instance, an \av that is spawned too close or even directly inside its goal area does not contribute to the question of whether the \av can safely drive.

Valid and interesting scenarios must meet the following conditions:

\noindent\textbf{Initial Position and Velocity}: The initial position of any \av is within the road network, and its initial velocity is between \minVelocity and \maxVelocity.

\noindent\textbf{Safety Distance}: There is enough distance between any pairs of autonomous vehicles (\av, $\av^{\prime}$) on the same lane to avoid a collision.

\noindent\textbf{Distance to Goal}: The initial position and goal area of any \av are at least \minDistGoal units apart (longitudinally).

\noindent\textbf{Goal Reachability}: The goal area of any \av is \emph{reachable} from its initial positions by navigating through the road network without violating any traffic rule.

\subsection{Fitness functions}
\label{sec:fitnessFunctions}
\approach executes a multi-AV scenario \scenario until a global execution timeout is reached or all the autonomous vehicles reach their goal area, collide, or drive off the road.
It collects the states \stateID{\av}{\timestamp_i} and planned trajectories $\plannedtrajectory^{\timestamp_{i}}_{\av}$ of each $\av \in \AVs$ at any timestamp \mbox{$\timestamp_i \in \{\timestamp_1,$ $\ldots,$ $\timestamp_n$\}} within the execution of the scenario.
Based on the collected data, \approach computes three fitness functions ({\it interaction fitness}, {\it interaction diversity fitness}, and {\it number of vehicles fitness}) that optimizes to generate scenarios in which \AVs interact frequently and in different ways while involving as few \AVs as possible.

\subsubsection{Interaction Fitness ($\nearrow$)}
\label{sec:fitnessInt}
The interaction fitness \fitnessInt{} accounts for the cumulative interactions observed during the execution of a scenario \scenario. 
\approach aims to maximize \fitnessInt{} to incentivize 
\begin{inparaenum}[(i)] 
\item critical decisions to avoid dangerous situations such as collisions, and 
\item strategic decisions to avoid violating traffic rules.
\end{inparaenum}
By doing so, \approach creates complex interplay among \AVs, aiming to increase the likelihood of exposing safety-critical issues in \AVs caused by emergent behaviors.

\noindent
We define \fitnessInt{} as:
\begin{equation}\label{eq:fitnessInt}
\fitnessInt(\scenario) = \sum_{\av \in \AVs} \left
|\{ \timestamp_{i} \in \{\timestamp_1, \ldots, \timestamp_n\} \colon \isInteraction(\av, \timestamp_{i}) \} \right|
\end{equation}
where $\isInteraction(\av,\timestamp_{i})$ = {\it true} if \av interacts with any other obstacle at time $\timestamp_{i}$ according to the definition we provided in Section~\ref{sec:interactions} and \scenario is a scenario that lasts $n$ time steps and involves $\AVs$ autonomous vehicles.

\subsubsection{Interaction Diversity Fitness ($\nearrow$)}
\label{sec:fitnessDiversity}
\approach generates scenarios that differ for the type of the interactions, which it quantifies in terms of changes of the trajectories of each \av for each interaction involving the vehicle \av at time $\timestamp_i$ (as described in Section~\ref{sec:interactions}).

The interaction diversity fitness \fitnessDiversity measures the differences between scenario \scenario and the previously computed scenarios, which \approach aims to maximize.
To compute the \fitnessDiversity achieved in a scenario \scenario, \approach records the type and number of interactions observed in \scenario in a vector \mbox{$\changeType_{\scenario} = [\# BL,$ $\#L,$ $\ldots,$ $\#R, \#FR]$}, and computes the minimum distance between $\changeType_{\scenario}$ and the vectors computed for the previously generated scenarios $\scenario_1$, \ldots, $\scenario_k$: 
\begin{equation}\label{eq:fitnessDiv}
\fitnessDiversity(\scenario) = \min_{\scenario_i \in \{\scenario_1, \ldots, \scenario_k\}} \| \changeType_{\scenario} - \changeType_{\scenario_i}\|_2
\end{equation}

\subsubsection{Number of Vehicles Fitness ($\searrow$)}
\label{sec:fitnessNumCars}
\approach limits the complexity of the generated scenarios by reducing the number of planning problems to the vehicles that (directly or indirectly) participate in the interactions.
Given a scenario \scenario with vehicles $\AVs$, \approach aims to minimize 
\begin{equation}\label{eq:fitnessNumCars}
\fitnessNumCars(\scenario) = |\AVs|
\end{equation}

\section{Evaluation Methodology}
\label{sec:evaluation}
This section presents the main research questions (Section~\ref{sec:research-questions}), the metrics  (Section~\ref{sec:evaluation-metrics}), and the experiments we executed to collect such metrics (Section~\ref{sec:experimental-setup}).

\subsection{Research Questions}
\label{sec:research-questions}
We answer the following research questions (RQs):

\newcommand{\rqOne}{Does \approach generate multiple and different types of interactions?\xspace}

\newcommand{\rqTwo}{Does \approach generate scenarios with multiple and different types of collisions?\xspace}

\newcommand{\rqThree}{Does \approach generate scenarios leading to traffic rule violations?\xspace}

\newcommand{\rqFour}{Does \approach generate simple scenarios?\xspace} 

\newcommand{\rqFive}{How many valid scenarios does \approach generate?\xspace}

\begin{resq}[(Interactions exposure): \rqOne]
This research question evaluates \approach's ability to generate scenarios involving multiple \AVs and resulting in many interactions of different types.
\end{resq}

\begin{resq}[(Collision exposure): \rqTwo]
This research question evaluates whether generating scenarios that expose a wide variety of interactions also reveal collisions, thereby confirming that covering interactions is central for revealing safety-critical situations and achieving better safety in autonomous vehicles.
\end{resq}

\begin{resq}[(Traffic rule violation exposure): \rqThree]
This research question complements the previous one by evaluating \approach's ability to expose additional \AVs misbehaviors related to traffic rule violations, strengthening the importance of covering interactions in \AVs testing.
\end{resq}

\begin{resq}[(Scenario complexity): \rqFour]
This research question checks the relevance of the generated scenarios to \AVs testing and debugging, in terms of the number of \AVs involved in the scenario, since scenarios that involve only a limited number of autonomous vehicles without reducing the complexity of the interactions, make it easy both testing and debugging.
\end{resq}

\begin{resq}[(Effectiveness of test generation): \rqFive]
This research question assesses the effectiveness of \approach. 
Addressing this research question is important for assessing the practical applicability of \approach, i.e., its ability to produce useful test cases in a manner comparable to day-to-day development practices.
\end{resq}

\subsection{Evaluation Metrics}
\label{sec:evaluation-metrics}
We define several metrics to experimentally answer the research questions. 
In the following, we present these metrics grouped by type.

\subsubsection{Interaction Metrics}
To address RQ1, we consider the total number of interactions (\totalInts) and the total number of unique interactions (\uniqueInts). 
We compute \totalInts as per Equation~\ref{eq:fitnessInt}, whereas we compute \uniqueInts using the interaction map described in Section~\ref{sec:Classifying-Interactions}. 
We build the interaction map using all the interactions exposed by the scenarios generated in a single run, and compute \uniqueInts as the number of cells in the interaction map that are covered by at least one interaction.

\subsubsection{Test Effectiveness Metrics (Test Oracles)}
A scenario might expose various issues in \AVs, including collisions and traffic rule violations.
To address RQ2, we consider the total number of found collisions (\totalColls) and the total number of unique collisions (\uniqueColls), whereas we consider the total number of times \AVs failed to handle traffic lights (\totalFailTrafficLights) and stop signs (\totalFailStop), as well as the total number of times \AVs violated the speed limits (\totalSpeeding) to address RQ3.
We compute the total number of collisions (\totalColls) by checking whether the bounding boxes of two vehicles intersect.
We compute the number of unique collisions \uniqueColls as the number of unique combinations of {\it point of impact} (\pointOfImpact), {\it relative impact velocity} (\relImpVel), and {\it relative impact angle} (\relImpAng) exposed in a run.

The point of impact $\pointOfImpact \in \{\frontLeft,$ $\frontRight,$ $\frontCenter,$ $\backLeft,$ $\backRight,$ $\backCenter,$ $\middleRight,$ $\middleLeft\}$ identifies which part of the \AVs is stroked when the collision occurs (\texttt{F}ront \texttt{L}eft/\texttt{R}ight/\texttt{C}enter, \texttt{B}ack \texttt{L}eft/\texttt{R}ight/\texttt{C}enter, \texttt{M}iddle \texttt{L}eft/\texttt{R}ight). 
The relative impact velocity \relImpVel is the relative velocity of the impact; we consider $11$ partitions of the relative velocity ($10$ partitions capture values between $0$ and $100$ Km/h, and one partition captures the edge case $>100$ Km/h).
The relative impact angle \relImpAng is the relative angle of impact. We consider $10$ partitions of the range $[0, 180]$ degrees and an additional partition for the remaining cases.
According to this classification, we identify $968$ ($8 \times 11 \times 11$) possible types of collisions combining $(\pointOfImpact, \relImpVel, \relImpAng)$.

We rely on the oracles defined by Huai et al.~\cite{doppleganger}, to account for traffic rule violations. 
We identify violations related to traffic lights and stop signs by cross-checking the recorded movement of each autonomous vehicle against the traffic lights and road signs' positions and recorded states, whereas we determine speeding violations by verifying whether the \av’s velocity exceeds the posted speed limit of the lane the \egovehicle is driving on.
We compute \totalFailTrafficLights, \totalFailStop, and \totalSpeeding by counting the number of oracle violations over the scenarios executed in a run.

\subsubsection{Scenario Complexity and Test Generation Metric}
To address RQ4, we consider the number of autonomous vehicles in each scenario (\nVehicles). 
To answer RQ5, we consider the total number of generated scenarios (\totalScenarios) and how many of them are valid (\validScenarios).

We compute \nVehicles as the total number of vehicles simulated in all executed scenarios. 
We compute \totalScenarios and \validScenarios following the guidelines for evaluating test generators of the SBFT~CPS~Testing~Tool~Competition~\cite{DBLP:conf/sbst/GambiJRZ22,SBFT2023CPScompReport,SBFT2024CPScompReport}: \totalScenarios accounts for all the generated scenarios, while \validScenarios accounts for the number of valid scenarios (see Section~\ref{sec:constraints}).

\subsection{Experimental Setup}
\label{sec:experimental-setup}
All experiments generate and validate a number of scenarios, and execute the valid scenarios in a \emph{simulator} against (multiple instances of) a \emph{test subject}, for a given \emph{generation budget} and \emph{road network}. 
The logic that generates the scenarios may use data collected during execution to generate additional scenarios in a feedback loop.

We executed experiments that cover a wide range of experimental settings, including six road networks (Section~\ref{sec:road-networks}) and two simulators and test subjects (Section~\ref{sec:test-subjects}), to increase the generality of our conclusions. 
The road networks represent highway, suburban, and urban maps; the simulators implement low- and high-fidelity simulations; and the test subjects adopt standard, yet fundamentally different, approaches to autonomous driving.

We follow the guidelines proposed by Arcuri and Briand~\cite{ArcuriICSE11}, to increase the robustness of our conclusions. 
We repeated all experiments multiple ($n=10$) times, established the significance of our conclusions using statistically appropriate tests (the Mann-Whitney U test with $\alpha = 0.05$ as the significance level), and assessed the effect size using Vargha and Delaney's \Atwelve.\footnote{To ease the results discussion, we map the \Atwelve values into labels, following Kitchenham et al.'s classification~\cite{kitchenham2017robust}: 
For $\Atwelve > 0.5$: {\it negligible difference} ($\Atwelve \in (0. 5, 0.556)$), {\it small difference} ($\Atwelve \in [0.556, 0.638)$), {\it medium difference} ($\Atwelve \in [0.638, 0.714)$), and {\it large difference} ($\Atwelve \ge 0.714$). We use a symmetric classification for $\Atwelve < 0.5$.}

We compare the results with those obtained with \doppelTest~\cite{doppleganger}, a state-of-the-art approach for generating multi-AV scenarios (Section~\ref{sec:baseline}), to contextualize the results achieved with \approach.
We evaluated each approach in its native execution environments and in the execution environment defined by the other approach, and configured both environments with the same execution budget and (as much as possible) the same hyperparameters (see Table~\ref {tab:confValues}).

\begin{table*}
\caption{Experimental evaluations configuration values}
\label{tab:confValues}
\renewcommand{\arraystretch}{1.25}
\resizebox{\linewidth}{!}{
\begin{tabular}{lrrp{180pt}}
\toprule
Parameter Name & \multicolumn{2}{c}{Value} & Explanation\\
\cmidrule{2-3}
& \frenetix & Apollo\\
\midrule
\simulationRetries & 3 hours & 24 hours & Generation budget measured in total time.\\
\initialPopulation & 10 & 10 & Size of the initial set of scenarios for the search algorithm. \\
\maxTimeStep & 200 & 300 & Maximum duration, i.e., timeout, for each scenario execution expressed in simulation steps.\\
\distToEnd & 10.0 & \notApplicable & Distance (in meters) of the goal area from the end of a lanelet. \\
\minAmountAVs; \maxAmountAVs & 2; 8 & 2; 5 & Minimum and maximum amount of autonomous vehicles in a scenario.\\
\minPositionModifier; \maxPositionModifier & -50; 50 & -50; 50 & Limit on the maximum variation for mutating \AVs' initial position along the lanelets in meters.\\
\minVelocity; \maxVelocity & 0; 25 & \notApplicable & Limit on the maximum variation for mutating \av's initial velocity in meters per second.\\
\minInitVelocity; \maxInitVelocity & 5; 10 & \notApplicable & Minimum and maximum values for setting \AVs' initial velocity in meters per second.\\
\emergencySlowDownAmount & 5 & \notApplicable & The maximum number of consecutive simulation steps in emergency mode before \AVs halt.\\
\minTrafficLightDuration & \notApplicable & 50 & Minimum duration of the green light of traffic lights.\\
\maxTrafficLightDuration & \notApplicable & 150 & Maximum duration of the green light of traffic lights.\\
\bottomrule
\end{tabular}}
\end{table*}

\subsubsection{Road Networks}
\label{sec:road-networks}
We selected six road networks that replicate real-world highways, suburban roads, and urban roads, ensuring they have the necessary elements to support basic driving maneuvers, such as keeping the lane, following vehicles ahead, overtaking and cutting in, merging into traffic, taking existing ramps, and navigating controlled intersections. 
We included suburban roads and highways as they are underrepresented in research~\cite{AVTesting-Survey}.

\begin{figure}
\centering
\begin{subfigure}{0.32\textwidth}
\centering
\includegraphics[width=\linewidth]{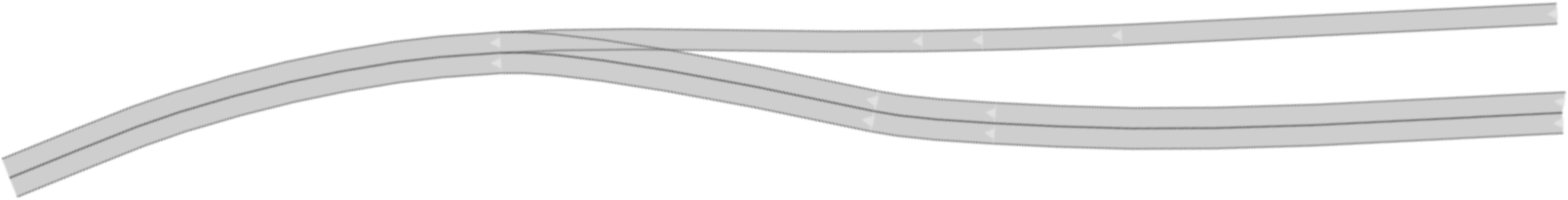}
\caption{\choID}
\label{fig:mergeScenario}
\end{subfigure}
%
\begin{subfigure}{0.32\textwidth}
\centering
\includegraphics[width=\linewidth]{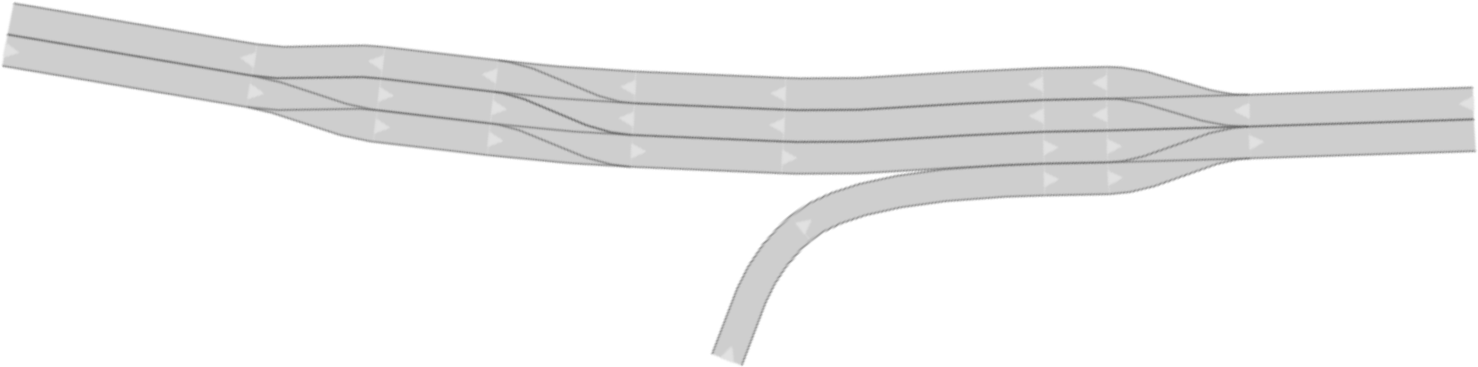}
\caption{\cologne}
\label{fig:oncomingLaneScenario}
\end{subfigure}
\\
\vspace{20pt}
\begin{subfigure}{0.32\textwidth}
\centering
\includegraphics[width=\linewidth]{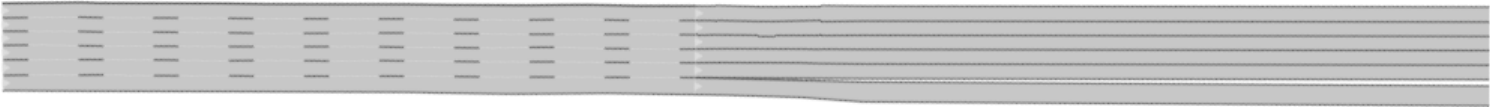}
\caption{\usaNine}
\label{fig:US101_09}
\end{subfigure}
\hfill
%
\begin{subfigure}{0.32\textwidth}
\centering
\includegraphics[width=\linewidth]{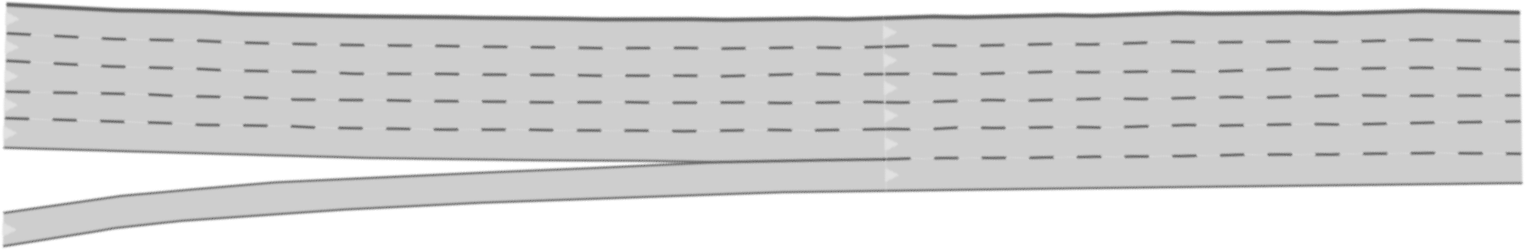}
\caption{\usaFifteen}
\label{fig:US101_15}
\end{subfigure}
\hfill
%
\begin{subfigure}{0.32\textwidth}
\centering
\includegraphics[width=\linewidth]{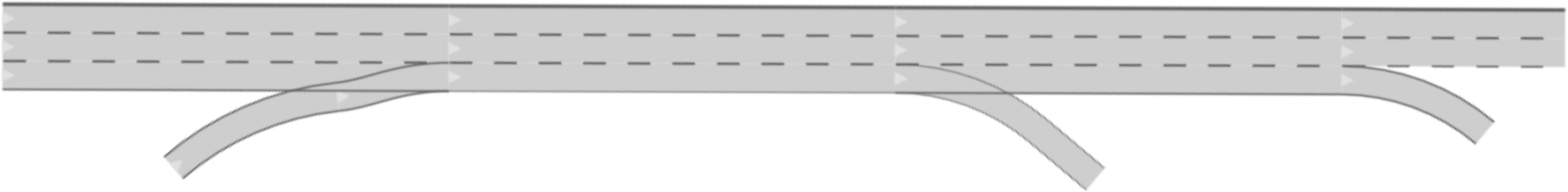}
\caption{\usaTwentythree}
\label{fig:US101_23}
\end{subfigure}
%
\caption{Benchmark road networks for test subject one (\frenetix)}
\label{fig:benchmarks-subject-one}
\end{figure}

We sampled the five road networks reported in Figure~\ref{fig:benchmarks-subject-one} from \CommonRoad's online library~\cite{commonRoadScenarios}.
Those road networks replicate segments of Asian (\choID) and European (\cologne) suburban roads as well as North American highways (\usaNine, \usaFifteen, and \usaTwentythree).

\choID is a long and almost straight single-lane road that merges into a two-lane road; thus, it forces the vehicles to tailgate each other, given the impossibility of overtaking in the narrow segment, and allows them to gain speed before merging. 
\cologne features lanes with opposite traffic directions, merges and bottlenecks, and an incoming lane that can lead to extreme situations such as head-on collisions.
\usaNine, \usaFifteen, and \usaTwentythree consist of multiple lanes, allow for high traffic flow and many opportunities for overtaking. 
They also feature entry and exit ramps that require \AVs to negotiate and synchronize their movement.

\begin{figure}
\centering
\begin{subfigure}{0.35\textwidth}
\centering
\includegraphics[width=\linewidth]{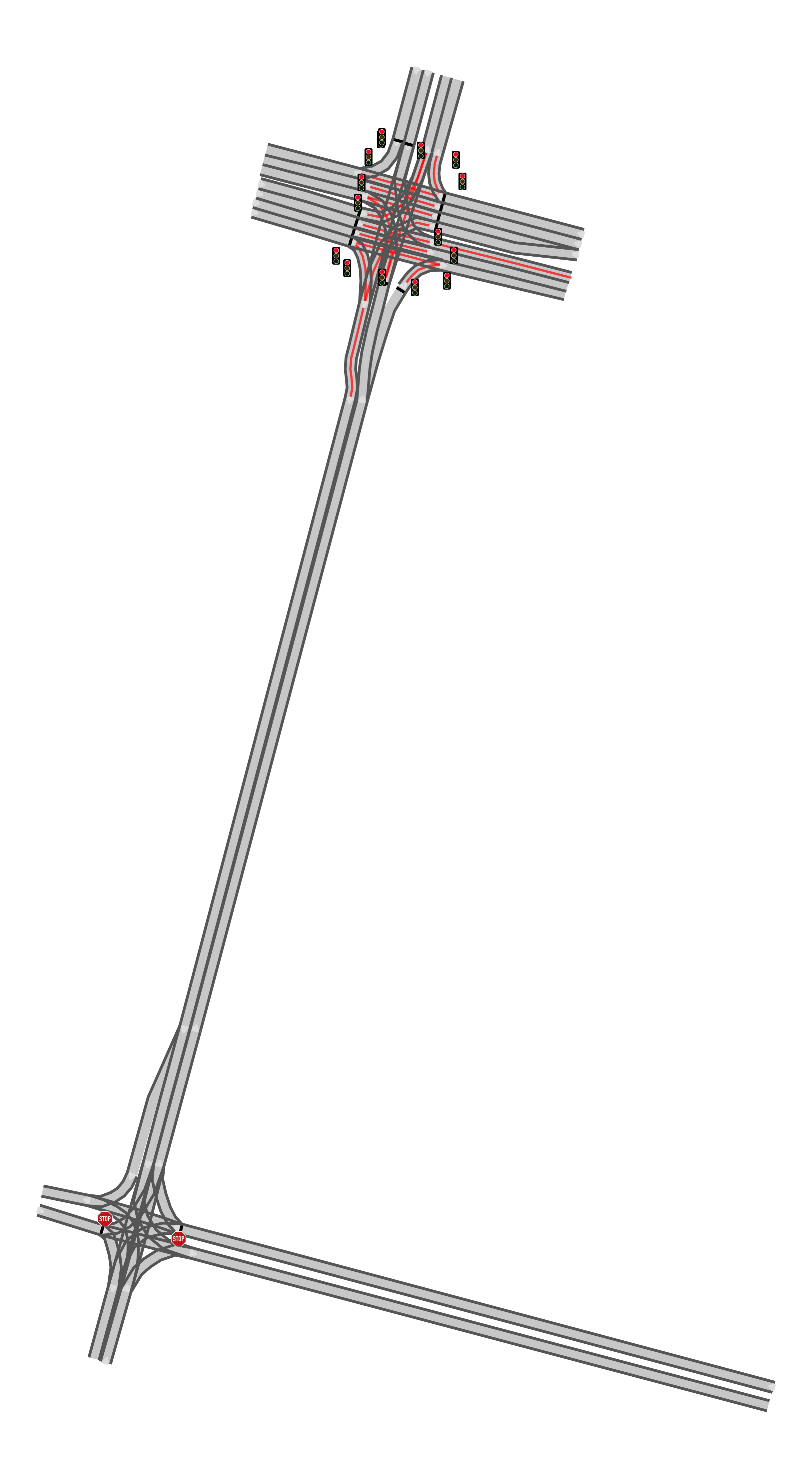}
\caption{\borregas}
\label{fig:borregasAve}
\end{subfigure}
%
%
\caption{Benchmark road network for test subject two (\baidu)}
\label{fig:benchmarks-subject-two}
\end{figure}

We replicated the experiments that Huai et al. executed to validate \doppelTest~\cite{replication-package-doppeltest} using a real-world urban street block in Sunnyvale, CA (\borregas), see Figure~\ref{fig:benchmarks-subject-two}.
\borregas includes a long straight segment connecting two four-way intersections, one controlled by traffic lights and one governed by stop signs.
This road network requires autonomous vehicles to handle traffic lights and negotiate the right-of-way at complex junctions and merging points.

\subsubsection{Test Subjects and Driving Simulators}
\label{sec:test-subjects}
We consider as test subjects two open-source \AVs implementations that have been extensively used in previous research. 
Each of these test subjects is designed to work with a specific simulator; therefore, in our evaluation, we consider two simulation environments: the \CommonRoad-based simulation framework proposed by Kaufeld et al.~\cite{multiagent2024} for \frenetix, and the \baidu-based simulation framework proposed by Huai et al.~\cite{doppleganger} for \apollo. 

\paragraph{FrenetiX}
\frenetix~\cite{Frenetix} is an efficient, open-source\footnote{\frenetix is available on GitHub at \url{https://github.com/TUM-AVS/Frenetix-Motion-Planner}} motion planner integrated with \CommonRoad.
This simulation platform combines a static global planner, called Route Planner, with a reactive local planner, called Behavioral Planner.
The Route Planner generates a reference path through the entire road network for the \av to reach the intended destination, i.e., the goal area.
The Behavioral Planner implements a sampling-based trajectory-planning cycle~\cite{DBLP:conf/icra/WerlingZKT10} to continuously plan trajectories that follow the reference path while optimizing safety (for instance, avoiding obstacles and driving only on allowed roads) and passenger comfort (for instance, avoiding abrupt accelerations).
\frenetix relies on a deep learning~\cite{DBLP:conf/smc/GeisslingerKBL21} component to predict the movement of nearby vehicles in the simulation.

\paragraph{\baidu}
\baidu\footnote{\baidu is available on GitHub at \url{https://github.com/ApolloAuto/apollo}} is an industry-level autonomous driving stack that adopts a modular architecture that integrates high-fidelity vehicle dynamics models, extensive sensing, and accurate control over ROS~\cite{ros}.
We use the version of \baidu available in the replication package of \doppelTest~\cite{replication-package-doppeltest}.

\subsubsection{Baseline Approach}
\label{sec:baseline}
We contextualize the achieved results by comparing \approach with \doppelTest~\cite{doppleganger}, the only state-of-the-art approach that, to the best of our knowledge, generates multi-AV critical scenarios.

Similarly to \approach, \doppelTest implements a multi-objective optimization algorithm to generate scenarios involving multiple \AVs in an urban context. 
Unlike \approach, \doppelTest also simulates pedestrians and directly uses the test oracles as fitness functions, an approach also known as \emph{falsification testing}~\cite{DBLP:journals/tecs/AbbasFSIG13,ARCHCOMP25Falsification}.
\doppelTest aims to trigger as many test oracles as possible, which has the downsides of not stressing a wide range of \AVs behaviors and of discovering the same issues multiple times, instead of generating scenarios that cover interactions.

\doppelTest original code works only for \apollo and its driving simulator, and it was tested only in urban scenarios. 
We extended \doppelTest to parse maps defined in \CommonRoad and compute its test oracles (see~Section~\ref{sec:evaluation-metrics}) from the simulation data collected by the multi-agent platform by Kaufeld et al.~\cite{multiagent2024}, to enable a fair comparison with \approach across different simulators and test subjects.
We make the code that extends \doppelTest freely available and open for review in our replication package~\cite{replication-package-evita}.
We implemented bindings to convert \apollo maps and simulations executed on \apollo simulator back into \CommonRoad, the format \approach uses to generate scenarios (see~Section~\ref{sec:background}) and evaluate their fitness (see~Section~\ref{sec:fitnessFunctions}). 
Our extensions enable the analysis and visualization of executed scenarios, thereby supporting post-processing, reporting, and human understanding.

\section{Experimental Results}
\label{sec:results}
This section summarizes the results we obtained by executing \approach and \doppelTest in the two experimental settings: against \frenetix, on the highway and suburban scenarios shown in Figure~\ref{fig:benchmarks-subject-one}, and \baidu, on the urban road layout shown in Figure~\ref{fig:benchmarks-subject-two}.
For all experiments, we used a Xeon Gold 6226R (16 core 2.9GHz) with 16GB of RAM.

\subsection{RQ1~--~\rqOne}
\label{sec:results-rq1}
RQ1 evaluates the ability of \approach to generate scenarios that trigger interactions, which we assess by counting the total number of interactions and the number of unique interactions.

\begin{figure}[!t]
\centering
\begin{subfigure}{0.675\linewidth}
\centering
\includegraphics[width=\linewidth]{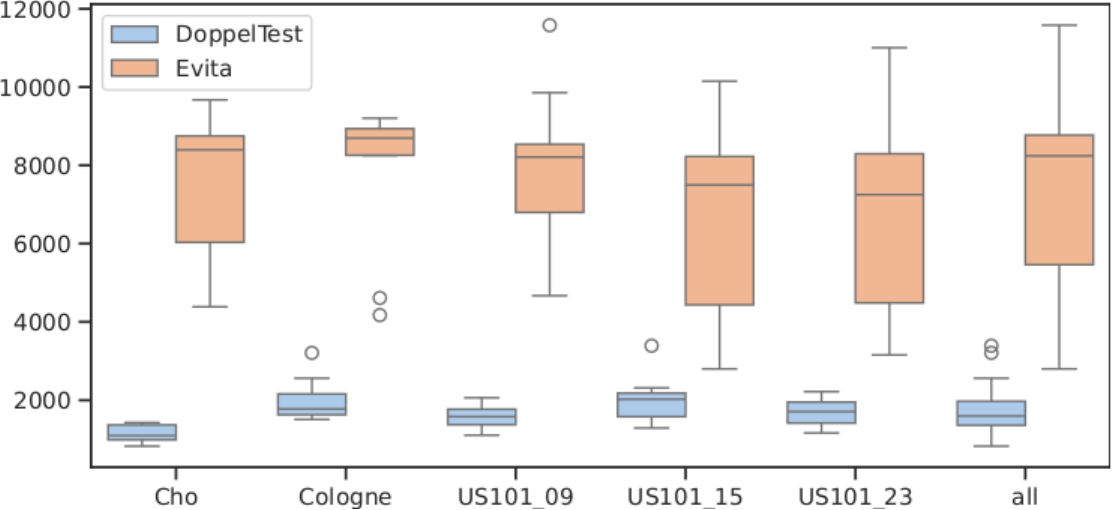}
\caption{Total number of interactions (\totalInts)}
\label{fig:rq1totalInteractions}
\end{subfigure}
\\
\vspace{10pt}
\begin{subfigure}{0.66\linewidth}
\centering
\includegraphics[width=\linewidth]{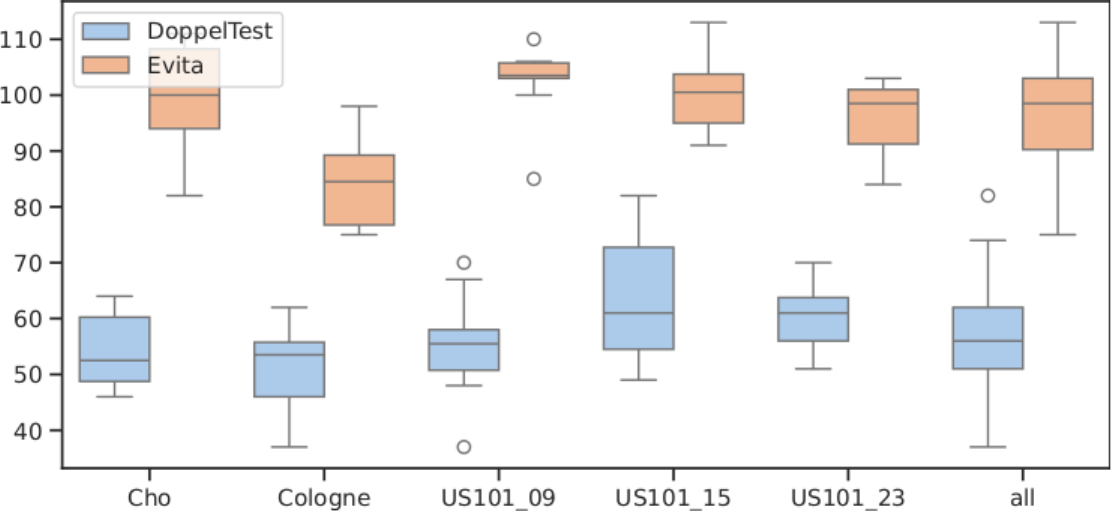}
\caption{Unique number of interactions (\uniqueInts)}
\label{fig:rq1uniqueInteractions}
\end{subfigure}
\caption{RQ1 -- Interactions found by \approach and \doppelTest with \frenetix experimental setting}
\label{fig:rq1interactions}
\end{figure}

\begin{figure}[!t]
\centering
\approach

\begin{subfigure}{0.19\textwidth}
\centering
\includegraphics[width=\linewidth]{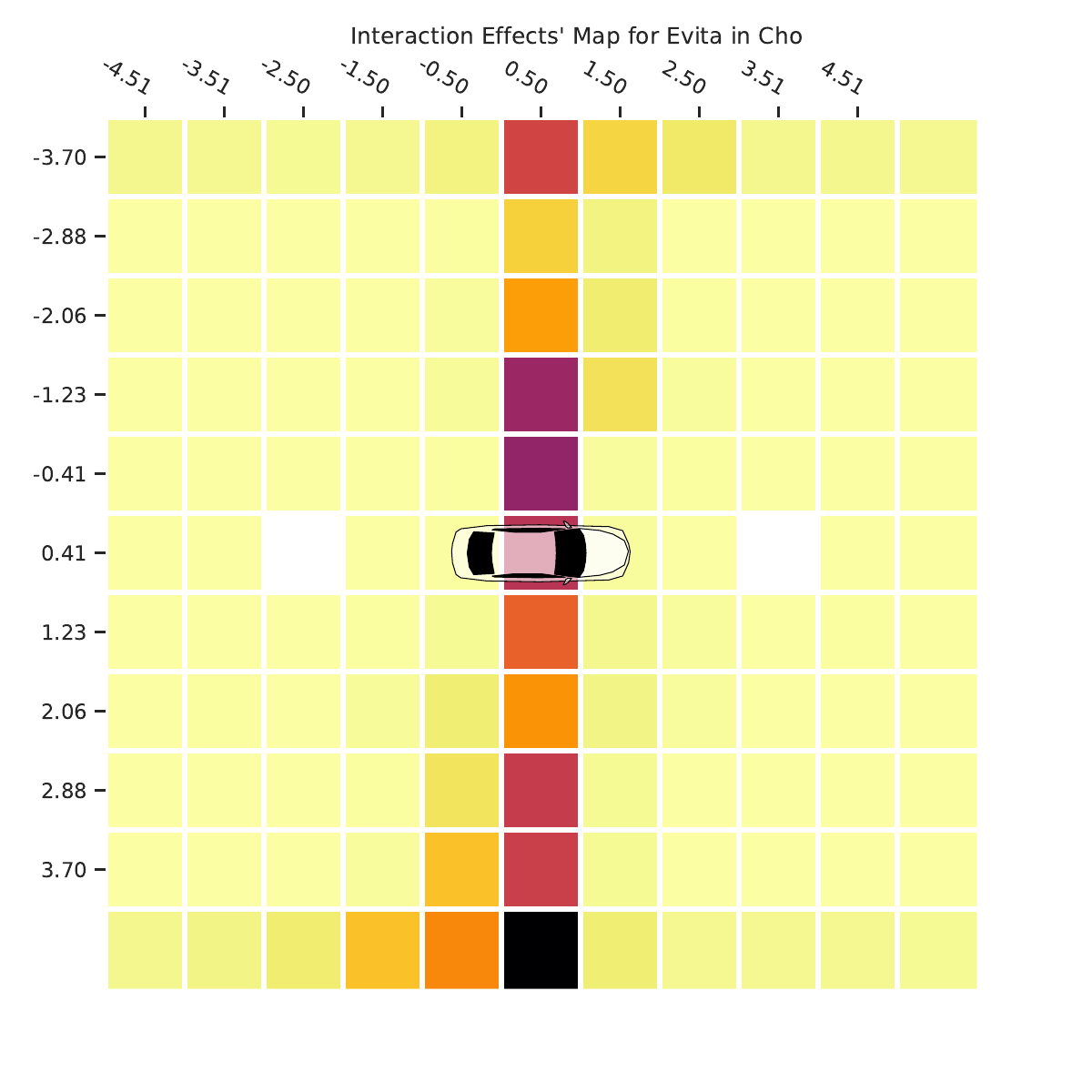}
\caption{\choID}
\label{fig:mergeScenarioIntMapEvita}
\end{subfigure}
\begin{subfigure}{0.19\textwidth}
\centering
\includegraphics[width=\linewidth]{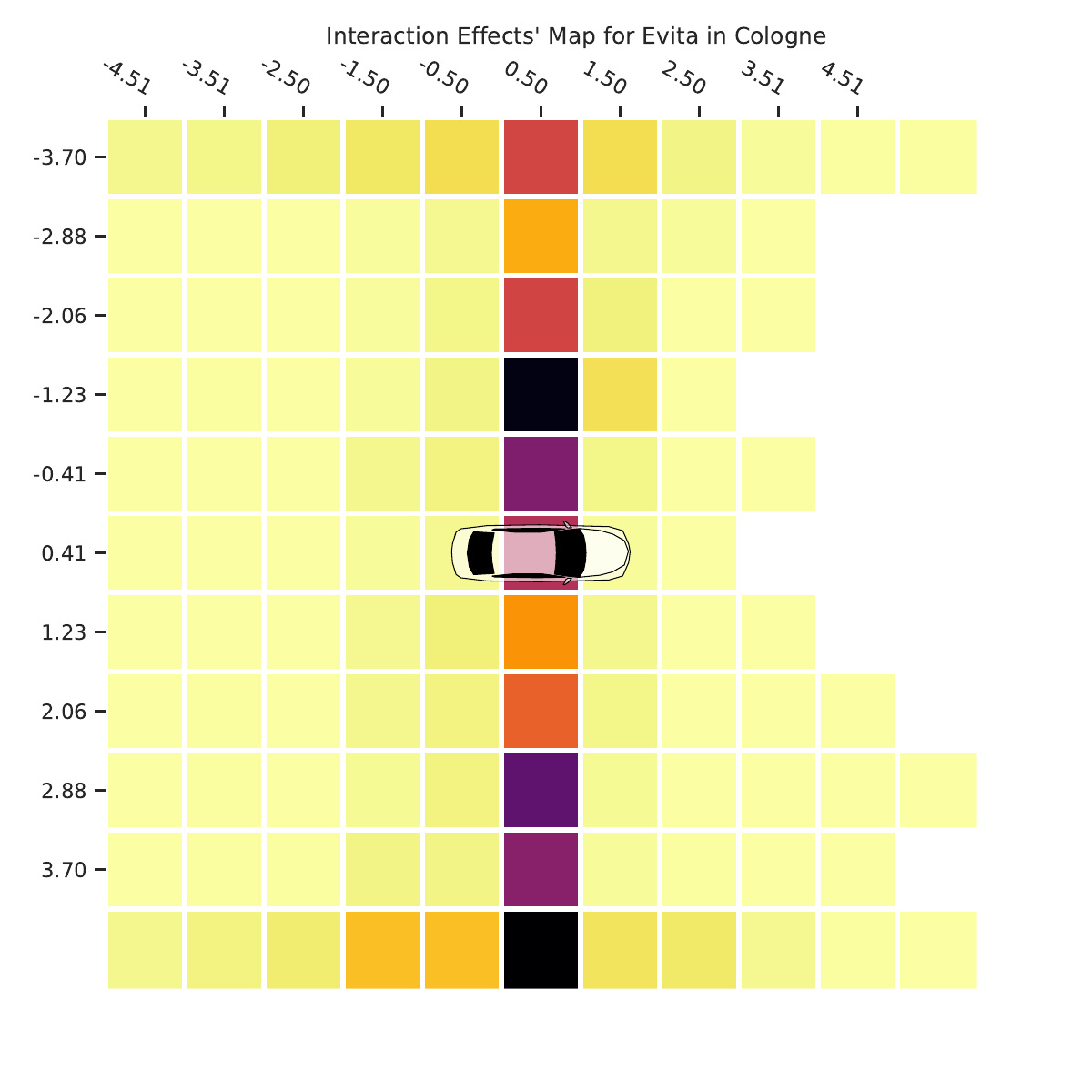}
\caption{\cologne}
\label{fig:oncomingLaneScenarioIntMapEvita}
\end{subfigure}
\begin{subfigure}{0.19\textwidth}
\centering
\includegraphics[width=\linewidth]{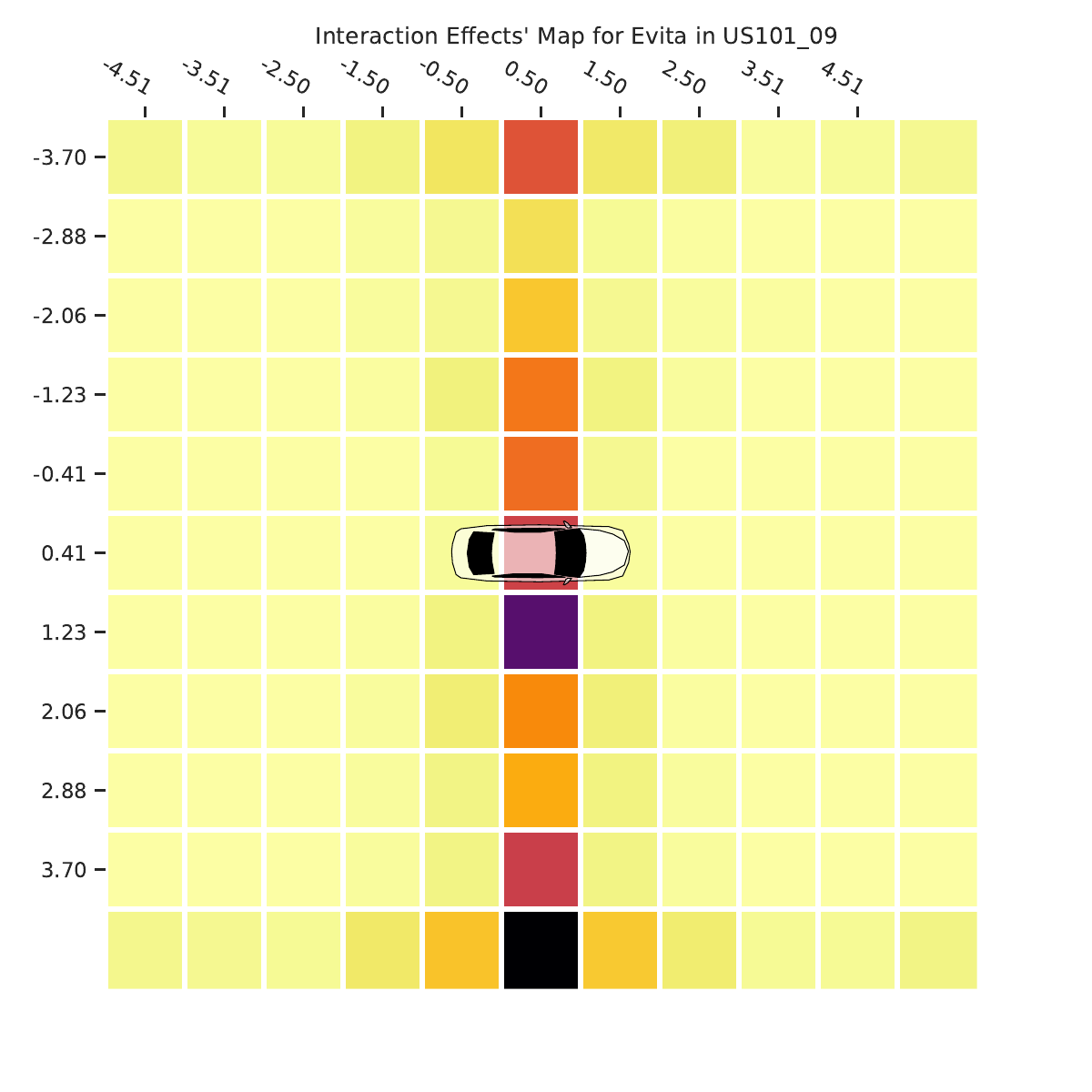}
\caption{\usaNine}
\label{fig:US101_09IntMapEvita}
\end{subfigure}
\begin{subfigure}{0.19\textwidth}
\centering
\includegraphics[width=\linewidth]{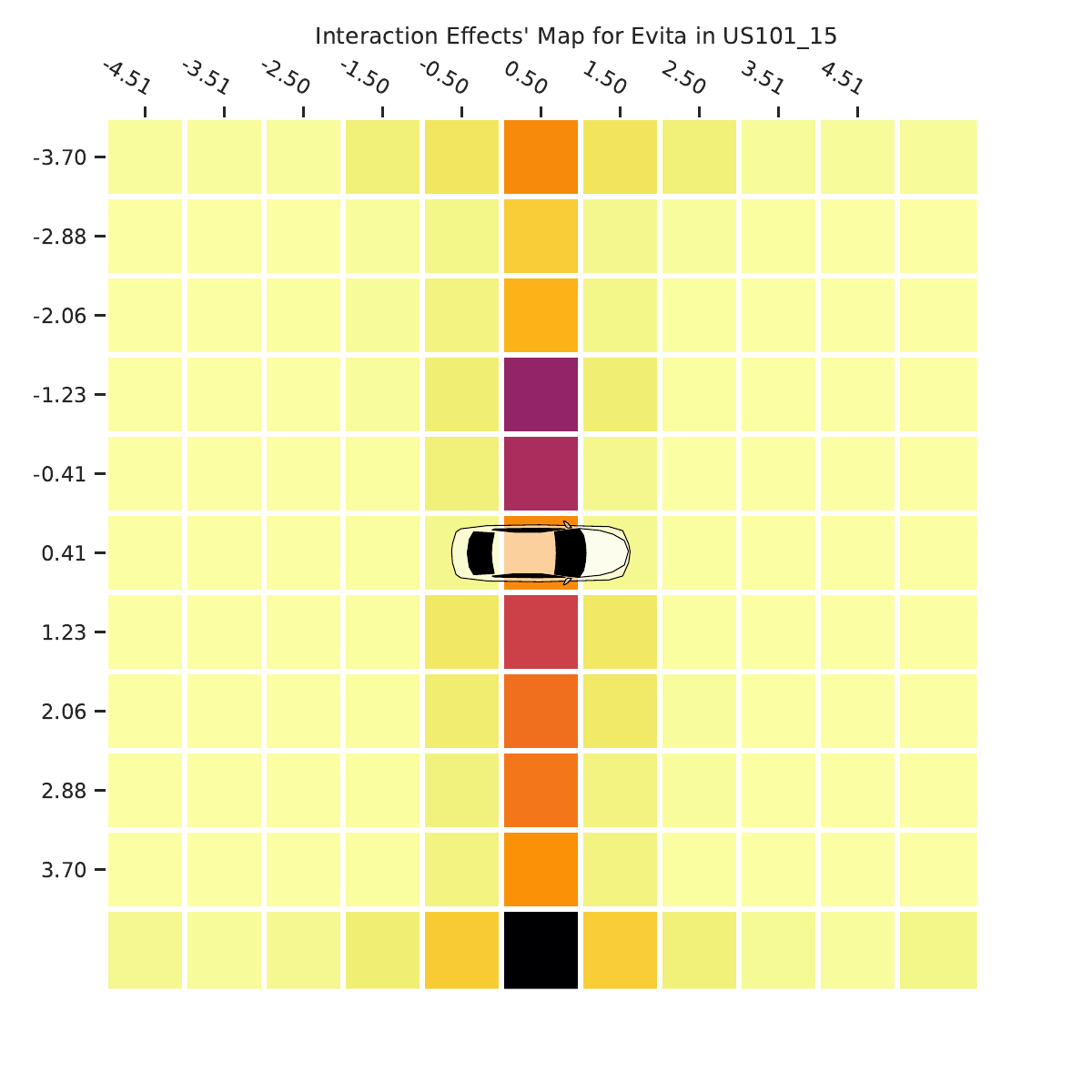}
\caption{\usaFifteen}
\label{fig:US101_15IntMapEvita}
\end{subfigure}
\begin{subfigure}{0.19\textwidth}
\centering
\includegraphics[width=\linewidth]{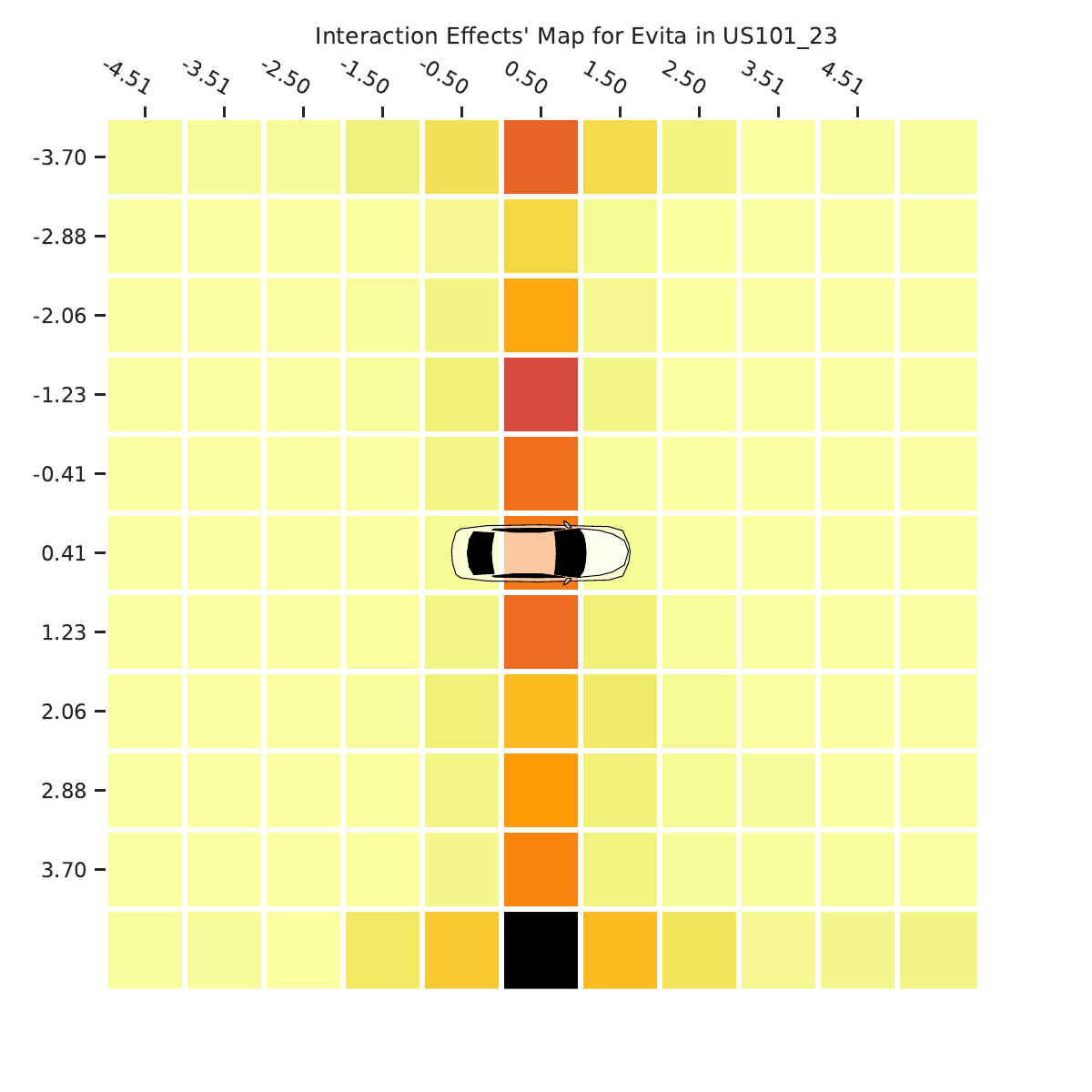}
\caption{\usaTwentythree}
\label{fig:US101_23IntMapEvita}
\end{subfigure}

\vspace{10pt}
\doppelTest

\begin{subfigure}{0.19\textwidth}
\centering
\includegraphics[width=\linewidth]{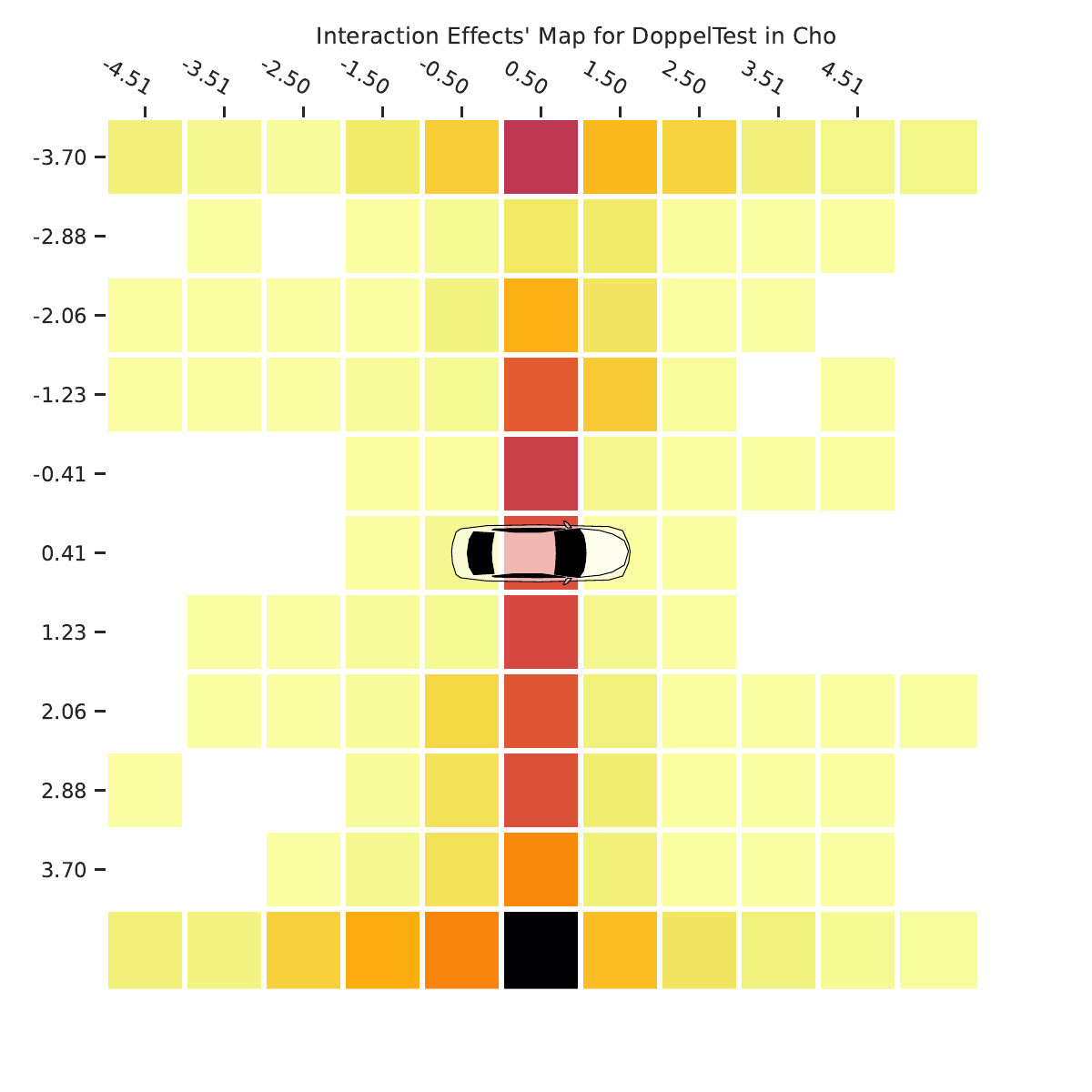}
\caption{\choID}
\label{fig:mergeScenarioIntMapDoppelTest}
\end{subfigure}
\begin{subfigure}{0.19\textwidth}
\centering
\includegraphics[width=\linewidth]{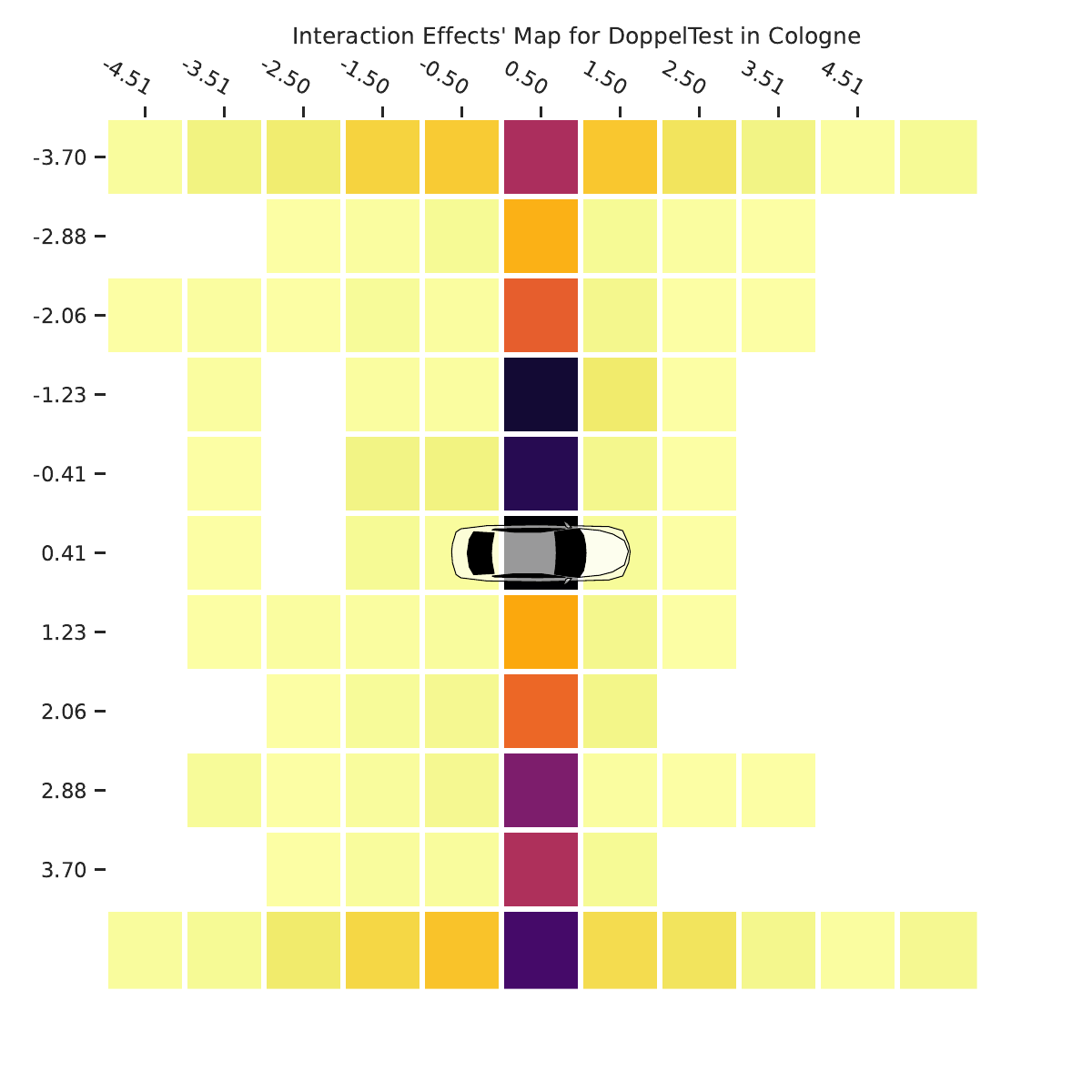}
\caption{\cologne}
\label{fig:oncomingLaneScenarioIntMapDoppelTest}
\end{subfigure}
\begin{subfigure}{0.19\textwidth}
\centering
\includegraphics[width=\linewidth]{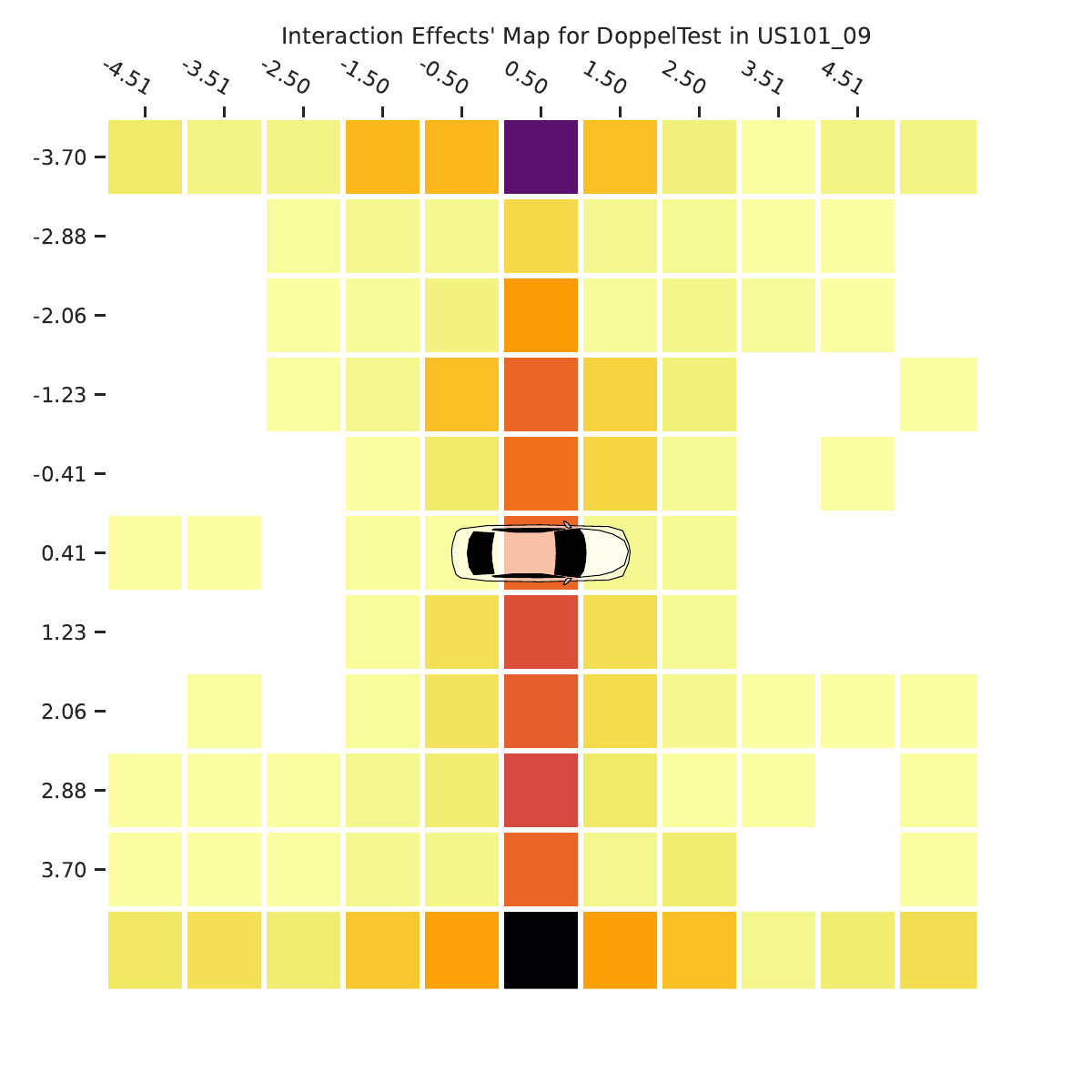}
\caption{\usaNine}
\label{fig:US101_09IntMapDoppelTest}
\end{subfigure}
\begin{subfigure}{0.19\textwidth}
\centering
\includegraphics[width=\linewidth]{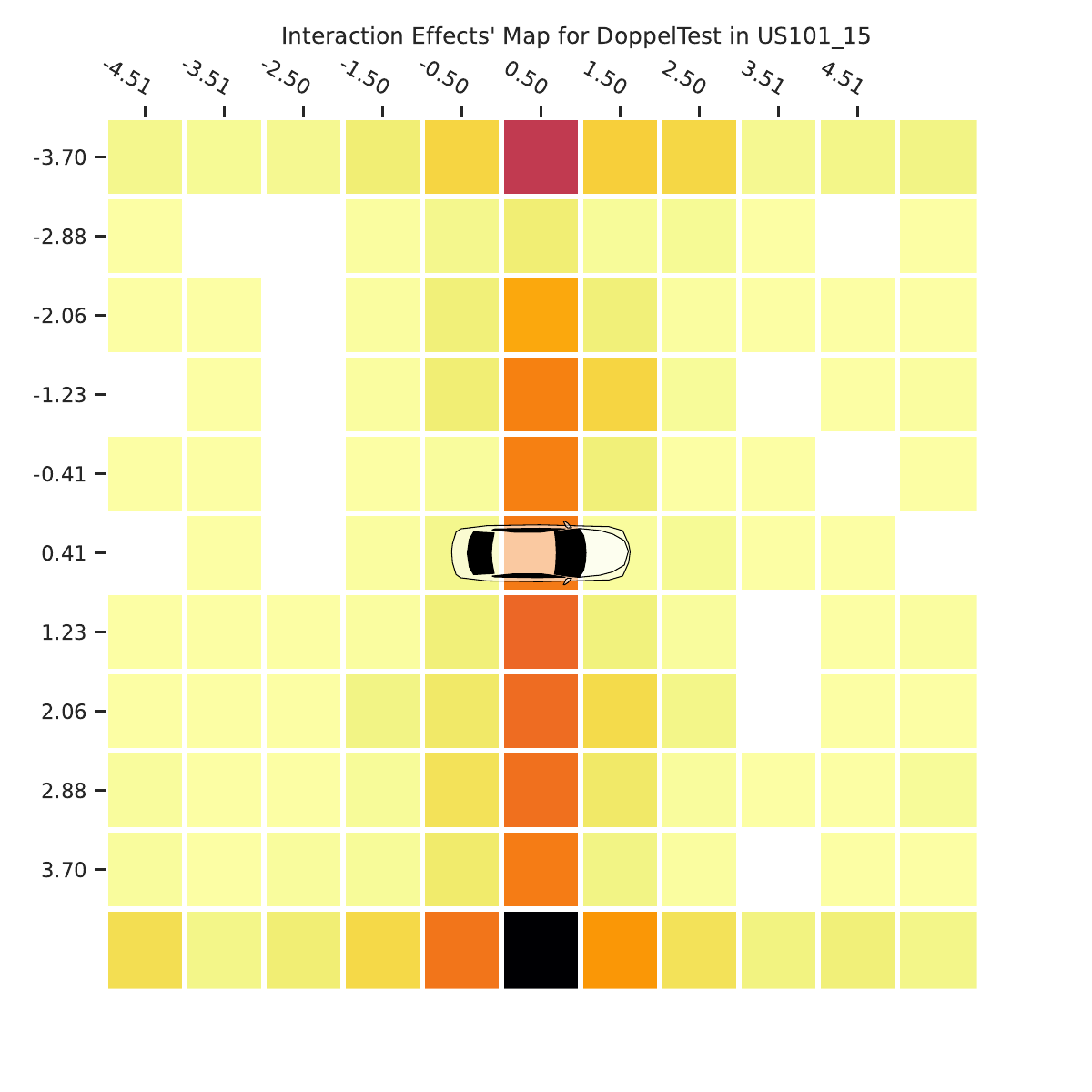}
\caption{\usaFifteen}
\label{fig:US101_15IntMapDoppelTest}
\end{subfigure}
\begin{subfigure}{0.19\textwidth}
\centering
\includegraphics[width=\linewidth]{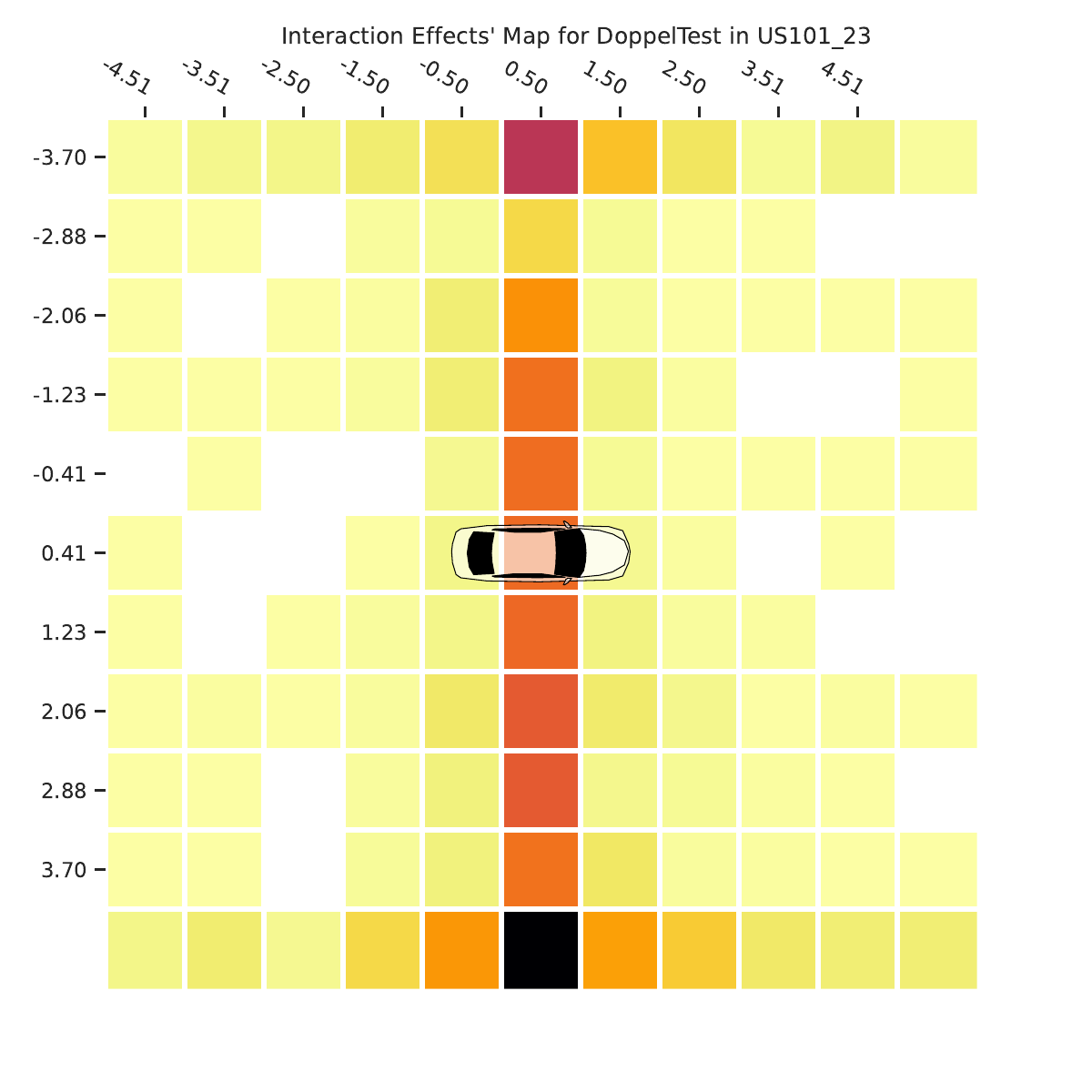}
\caption{\usaTwentythree}
\label{fig:US101_23IntMapDoppelTest}
\end{subfigure}
\caption{RQ1 -- Interactions maps with \frenetix experimental setting}
\Description{RQ1 -- Interactions maps with \frenetix experimental setting}
\label{fig:rq1interactionMaps}
\end{figure}

Figures~\ref{fig:rq1totalInteractions} and~\ref{fig:rq1uniqueInteractions} report the distribution of \totalInts and \uniqueInts obtained with \approach and \doppelTest across all the lane networks used in the experimental setting with \frenetix.
The right-most group in each plot shows the aggregated values across all the road networks.
Figure~\ref{fig:rq1interactionMaps} reports the interaction maps for each road network and approach.
Figure~\ref{fig:rq1interactions} clearly indicates that \approach always generates more interactions. The statistical analysis confirms the superiority of \approach on \doppelTest, as \approach is always statistically significantly better than \doppelTest in terms of \totalInts, with a large effect size. 
The interaction maps reported in Figure~\ref{fig:rq1interactionMaps} show that \approach covered more types of interactions than \doppelTest, as \approach's interaction maps contain fewer empty (white) cells.
This observation not only strengthens our conclusion that \approach generates more unique interactions than \doppelTest but also that \approach covers many, sometimes all, possible types of interactions when applied to suburban and highway road networks.

\medskip

\begin{figure}[!t]
\centering
\begin{subfigure}{0.49\linewidth}
\centering
\includegraphics[width = \linewidth]{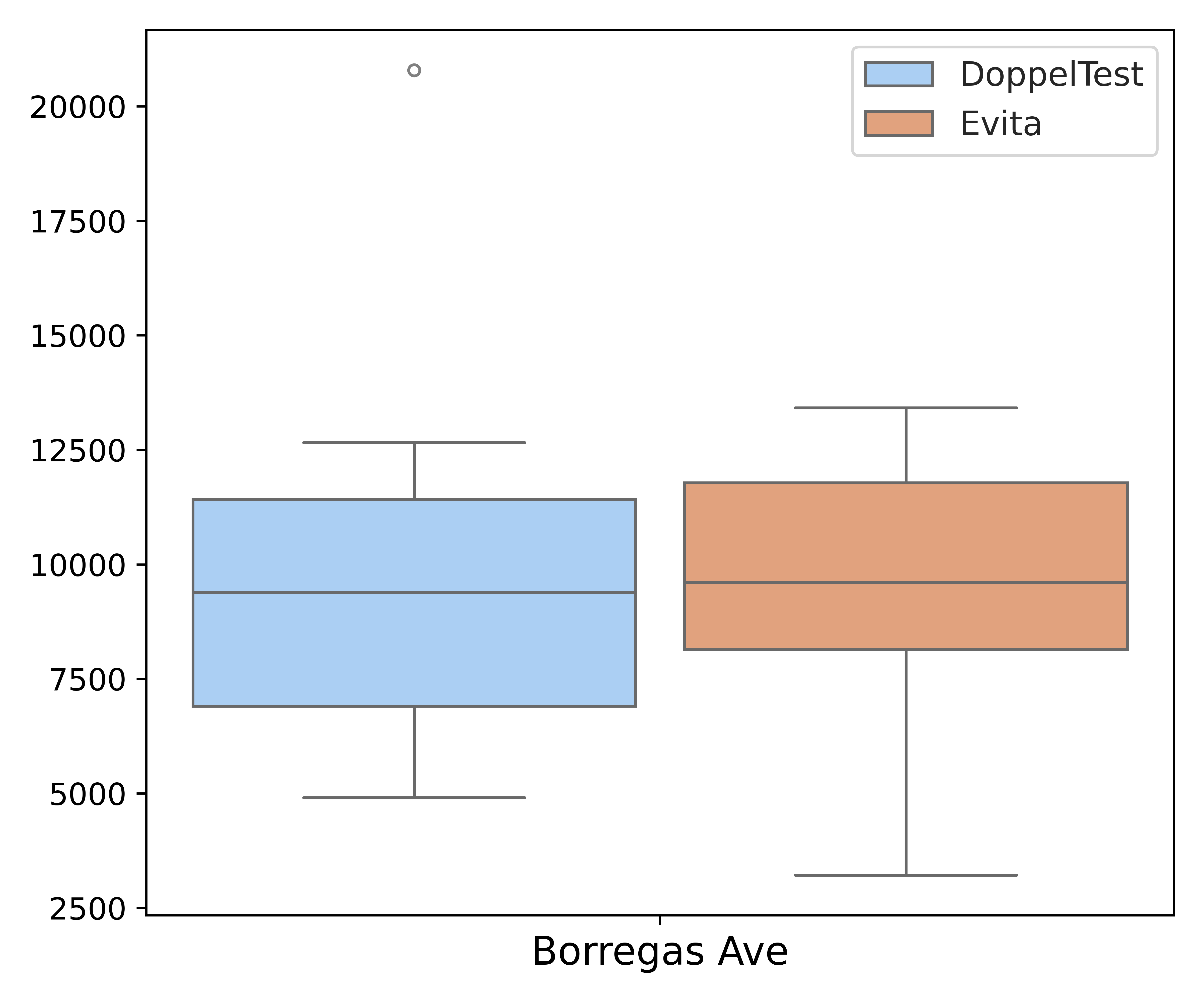}
\caption{Total number of interactions (\totalInts)}
\label{fig:rq1totalInteractionsBorregas}
\end{subfigure}
\begin{subfigure}{0.49\linewidth}
\centering
\includegraphics[width = \linewidth]{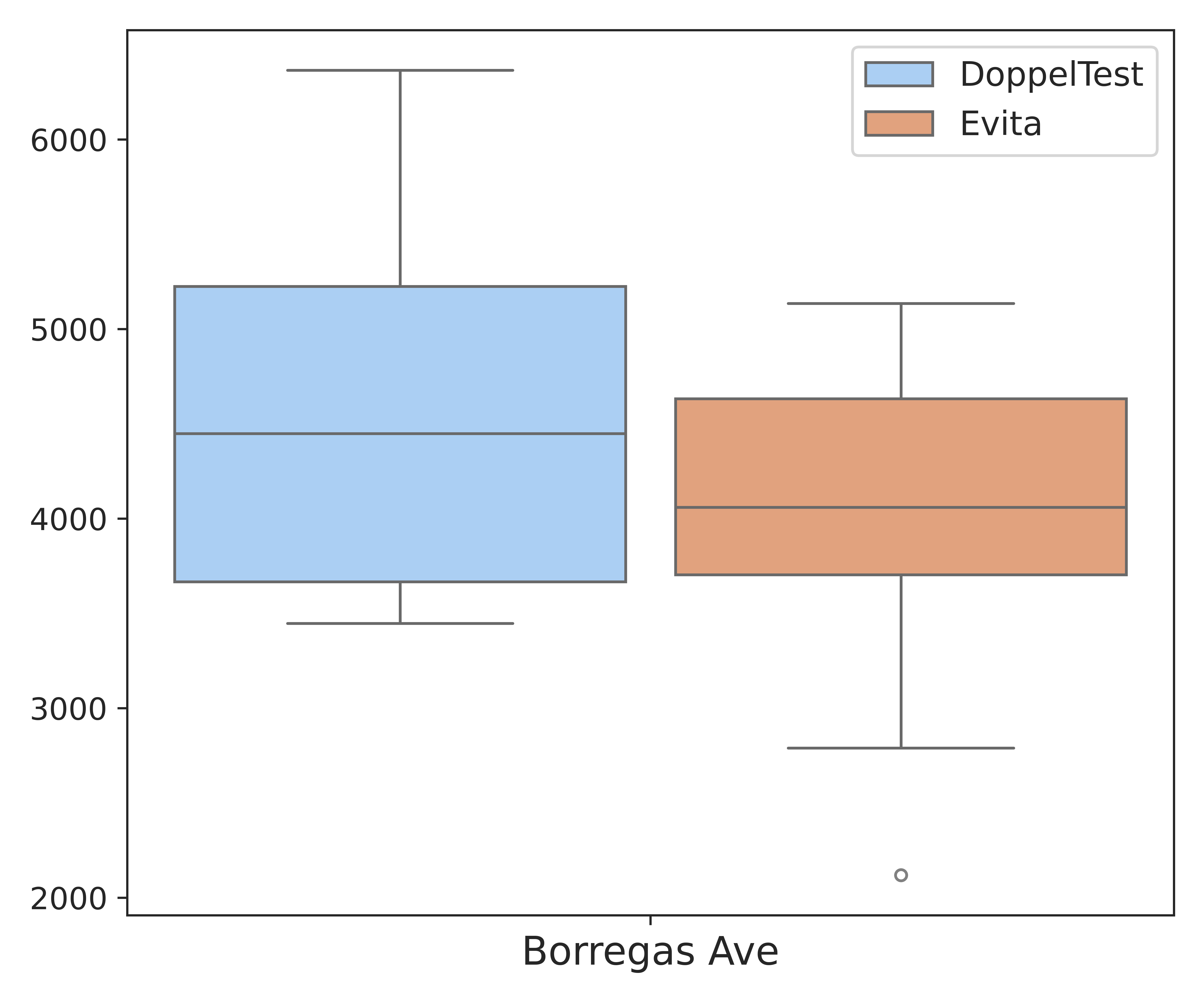}
\caption{Unique number of interactions (\uniqueInts)}
\label{fig:rq1uniqueInteractionsBorregas}
\end{subfigure}
\caption{RQ1 -- Interactions found by \approach and \doppelTest with \baidu experimental setting}
\label{fig:rq1interactionsBorregas}
\end{figure}

\begin{figure}[!t]
\centering
\begin{subfigure}{0.4\linewidth}
\centering
\includegraphics[width = 0.7\linewidth]{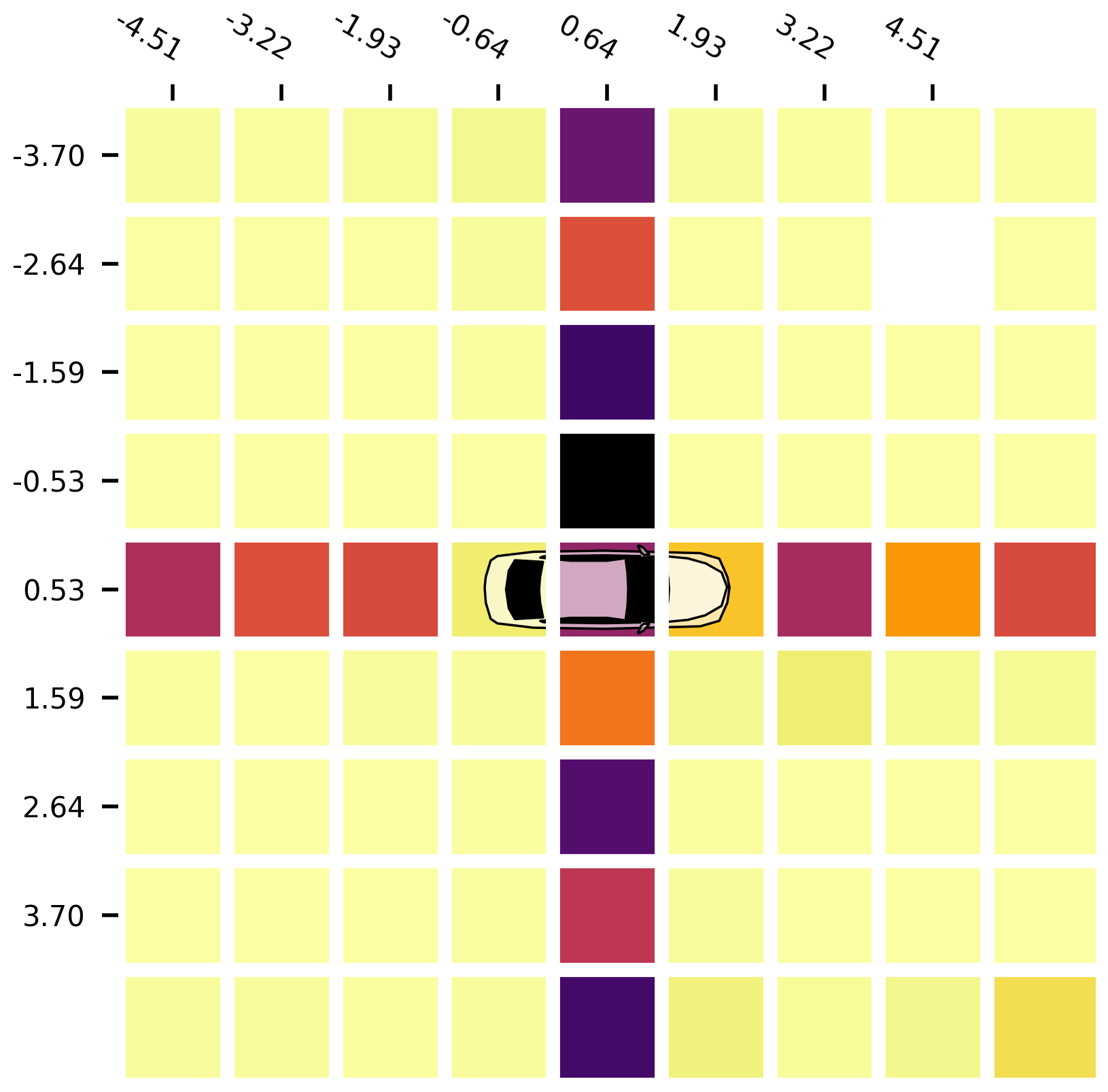}
\caption{\approach}
\label{fig:rq1EVITAInteractionMapBorregas}
\end{subfigure}
\begin{subfigure}{0.4\linewidth}
\centering
\includegraphics[width = 0.7\linewidth]{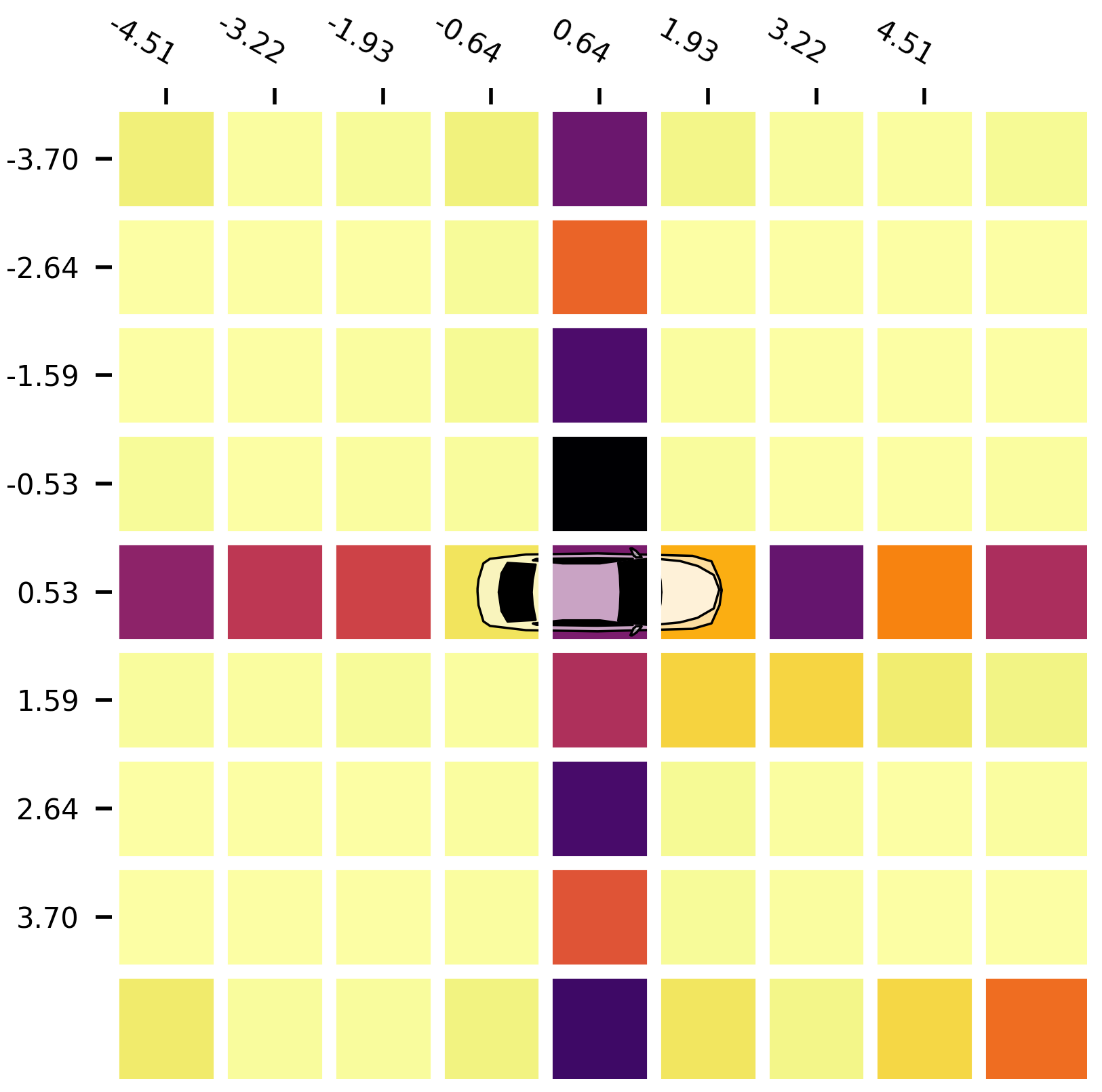}
\caption{\doppelTest}
\label{fig:rq1DoppelTestInteractionMapBorregas}
\end{subfigure}
\caption{RQ1 -- Interactions maps with \baidu experimental setting}
\label{fig:rq1interactionMapsBorregas}
\end{figure}

Figure~\ref{fig:rq1totalInteractionsBorregas} and Figure~\ref{fig:rq1uniqueInteractionsBorregas} report the distribution of \totalInts and \uniqueInts obtained with \approach and \doppelTest across the urban road network used in the experimental setting with \baidu.
Figure~\ref{fig:rq1interactionMapsBorregas} reports the interaction maps for the selected urban road network and the two approaches.

Figure~\ref{fig:rq1totalInteractionsBorregas} indicates that \approach and \doppelTest trigger a similar amount of interactions (\totalInts) and unique interactions (\uniqueInts) on the \borregas map, as confirmed by the statistical analysis.
The distribution of \uniqueInts represented in Figure~\ref{fig:rq1uniqueInteractionsBorregas} and the interaction maps in Figure~\ref{fig:rq1interactionMapsBorregas} indicate a small statistical disadvantage of \approach in the overall variety of the triggered interactions.
We argue that this marginal superiority of \doppelTest may be due to a higher complexity of the generated scenarios that heavily rely on a larger number of vehicles.
While \doppelTest can cover few additional unique interaction types, \approach remains highly competitive maintaining a lower scenario complexity.
We will further discuss the relationship between framework efficiency and scenario complexity in RQ4 (Section~\ref{sec:results-rq4}).

\begin{tcolorbox}[size = title, colframe = white, width = 1\linewidth, colback = gray!20, breakable]
{\bf Answer to RQ1.} \approach finds more and more diverse interactions than \doppelTest.
While \doppelTest shows a small statistical advantage in variety under specific urban settings, it requires scenarios with a significantly higher complexity, whereas \approach consistently covers a broad, and sometimes full, spectrum of interaction types across a wide array of road networks with simpler scenarios.
\end{tcolorbox}

\begin{remark}
The interaction maps in Figure~\ref{fig:rq1interactionMaps} highlight some interesting patterns:
Most interactions across all scenarios cluster around the central vertical line of the interaction maps, meaning that most interactions caused changes in \AVs' direction (for instance, lane switching) but not speed. 
We argue that this behavior relates to how \frenetix plans the trajectories to improve safety (by changing lanes to avoid obstacles) while minimizing passenger discomfort (by avoiding drastic speed changes).
\end{remark}

\subsection{RQ2~--~\rqTwo}
\label{sec:results-rq2}
RQ2 evaluates the ability to generate critical scenarios and compares \approach to \doppelTest in terms of both the number and types of triggered collisions.

\begin{figure}[!t]
\centering
\begin{subfigure}{0.675\linewidth}
\centering
\includegraphics[width=\linewidth]{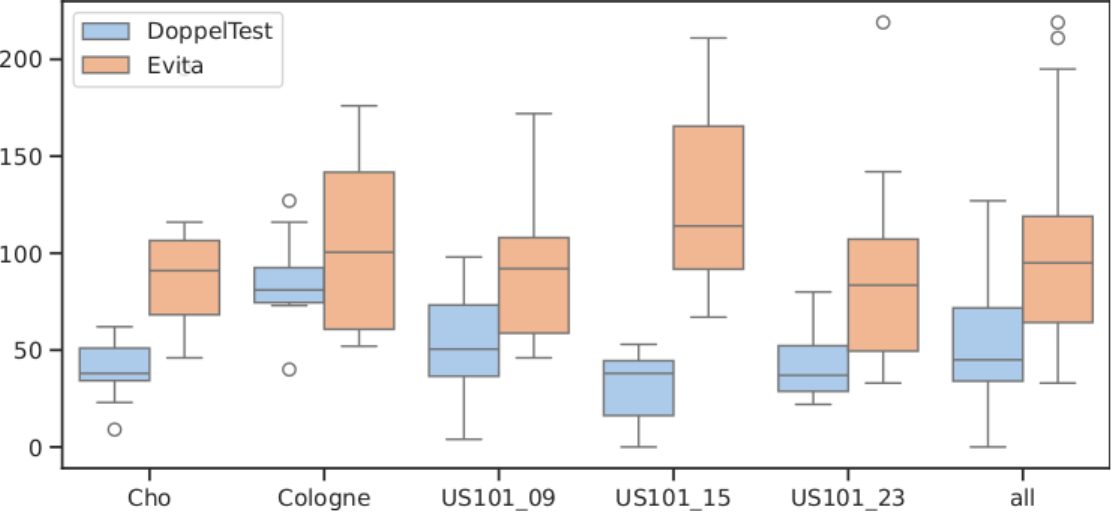}
\caption{Total number of collisions (\totalColls)}
\label{fig:rq2totalCollisions}
\end{subfigure}
\\
\vspace{10pt}
\begin{subfigure}{0.67\linewidth}
\centering
\includegraphics[width=\linewidth]{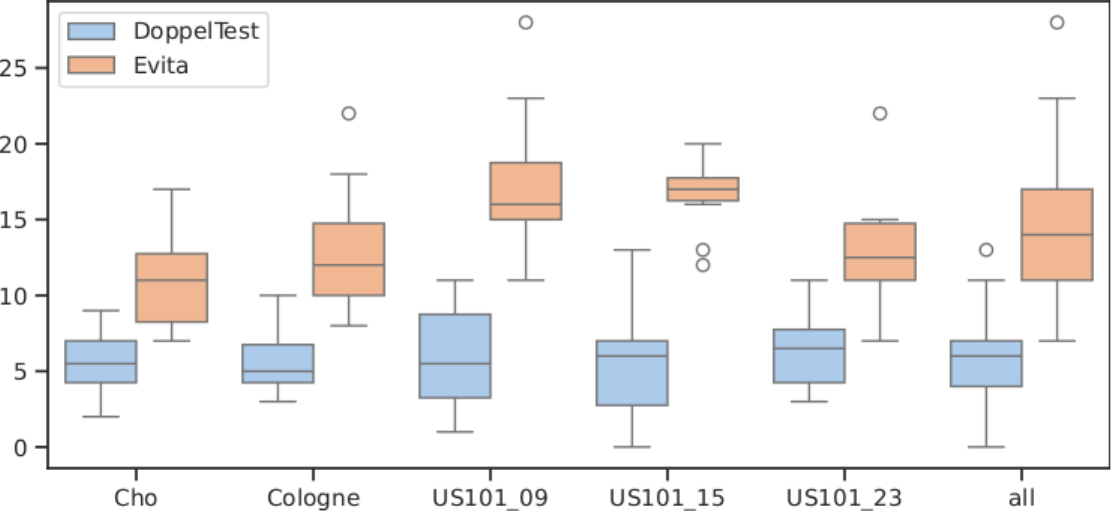}
\caption{Unique number of collisions (\uniqueColls)}
\label{fig:rq2uniqueCollisions}
\end{subfigure}
\caption{RQ2 -- Collisions found by \approach and \doppelTest with \frenetix experimental setting}
\label{fig:rq2collisions}
\end{figure}

Figure~\ref{fig:rq2totalCollisions} reports the total number of \approach and \doppelTest scenarios that lead to collisions for the different road networks and in total (rightmost plot) in the \frenetix experimental setting.
Figure~\ref{fig:rq2uniqueCollisions} reports the number of unique collisions \uniqueColls. 

Figure~\ref{fig:rq2uniqueCollisions} shows that \approach is always statistically significantly better than \doppelTest in terms of \uniqueColls, with a large effect size.
Figure~\ref{fig:rq2totalCollisions} shows the statistical advantage of \approach over \doppelTest in terms of \totalColls for all road network but \cologne, for which there is no statistically significant difference.

\medskip

In the \baidu experimental setting \approach generates \TotalCollsBorregasEvita collisions and \UniqueCollsBorregasEvita unique collisions, while \doppelTest generates \TotalCollsBorregasDopple collisions and \UniqueCollsBorregasDopple unique collisions, with a similar number of generated interactions.
The data confirm the superiority of the \approach in generating relevant scenarios also in urban road networks.

\medskip

As discussed in Section~\ref{sec:evaluation-metrics}, \uniqueColls classify the collisions in terms of point of impact (\pointOfImpact), relative velocity of the impact (\relImpVel), and relative angle of the impact (\relImpAng).

\paragraph{Point of impact}
\approach covers all the types of \pointOfImpact, and finds more collisions than \doppelTest for most of the types, in both highway and suburban road networks.
Both \approach and \doppelTest generate many front collisions, few back collisions, and rarely middle collisions.
Although middle collisions are quite rare for both \approach and \doppelTest, \approach finds more middle collisions than \doppelTest.
In urban scenarios, the majority of collisions generated by the two frameworks are middle collisions, followed by front collisions, while back collisions are quite rare.

\paragraph{Relative impact speed}
Both \approach and \doppelTest collisions cluster around $10$~km/h and $20$~km/h, for both the experimental settings. 
\approach almost always finds more collisions than \doppelTest in both the $10$~km/h and $20$~km/h clusters.

\paragraph{Relative impact angle}
All collisions fall within the $20$~Deg cluster in both \frenetix and \baidu experimental settings and correspond to failed overtaking.
We report exemplary cases in the replication package~\cite{replication-package-evita}.
Despite collisions usually occurring on straight road segments, \approach covers a wider range of relative impact angles than \doppelTest.

\begin{tcolorbox}[size=title, colframe=white, width=1\linewidth, colback=gray!20
, breakable
]
{\bf Answer to RQ2.} \approach finds both more and more diverse collisions than \doppelTest.
\end{tcolorbox}

\begin{remark}
The results of RQ1 and RQ2 confirm our intuition that the more diverse are the interactions, the more diverse are the collisions.
\end{remark}

\subsection{RQ3~--~\rqThree}
\label{sec:results-rq3}
We address RQ3 with experiments on the \borregas urban road layout of \baidu, since the selected maps of the highway and suburban road networks of \frenetix do not include traffic lights and signs.
We refer to Huai~et~al.'s~\cite{doppleganger} definition to identify traffic violations. 

\begin{figure}[!t]
\centering
\begin{subfigure}{0.4\textwidth}
\centering
\includegraphics[width=\linewidth]{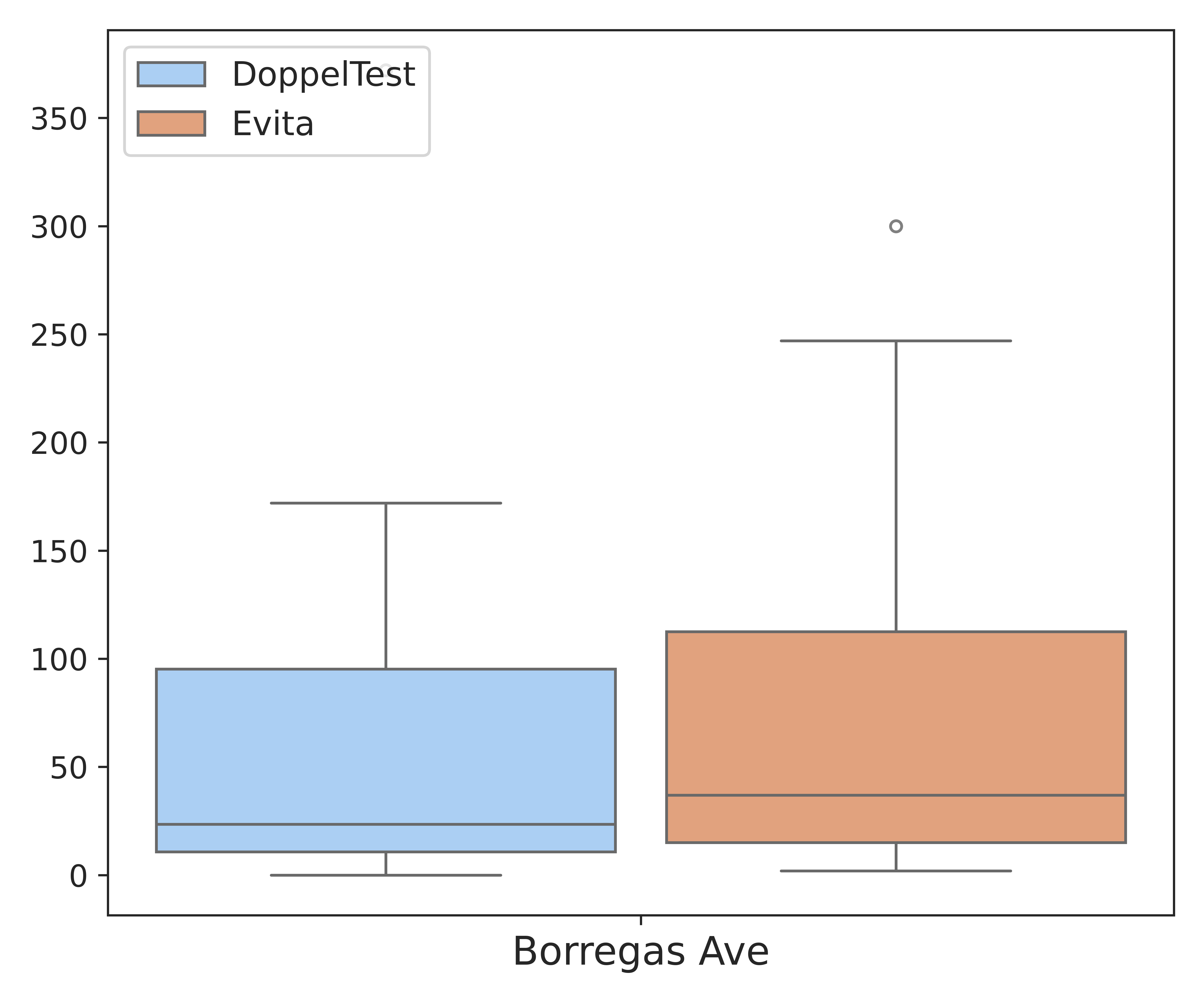}
\caption{Stop Signs Violations}
\label{fig:stopSignsViolations}
\end{subfigure}
\qquad
%
\begin{subfigure}{0.4\textwidth}
\centering
\includegraphics[width=\linewidth]{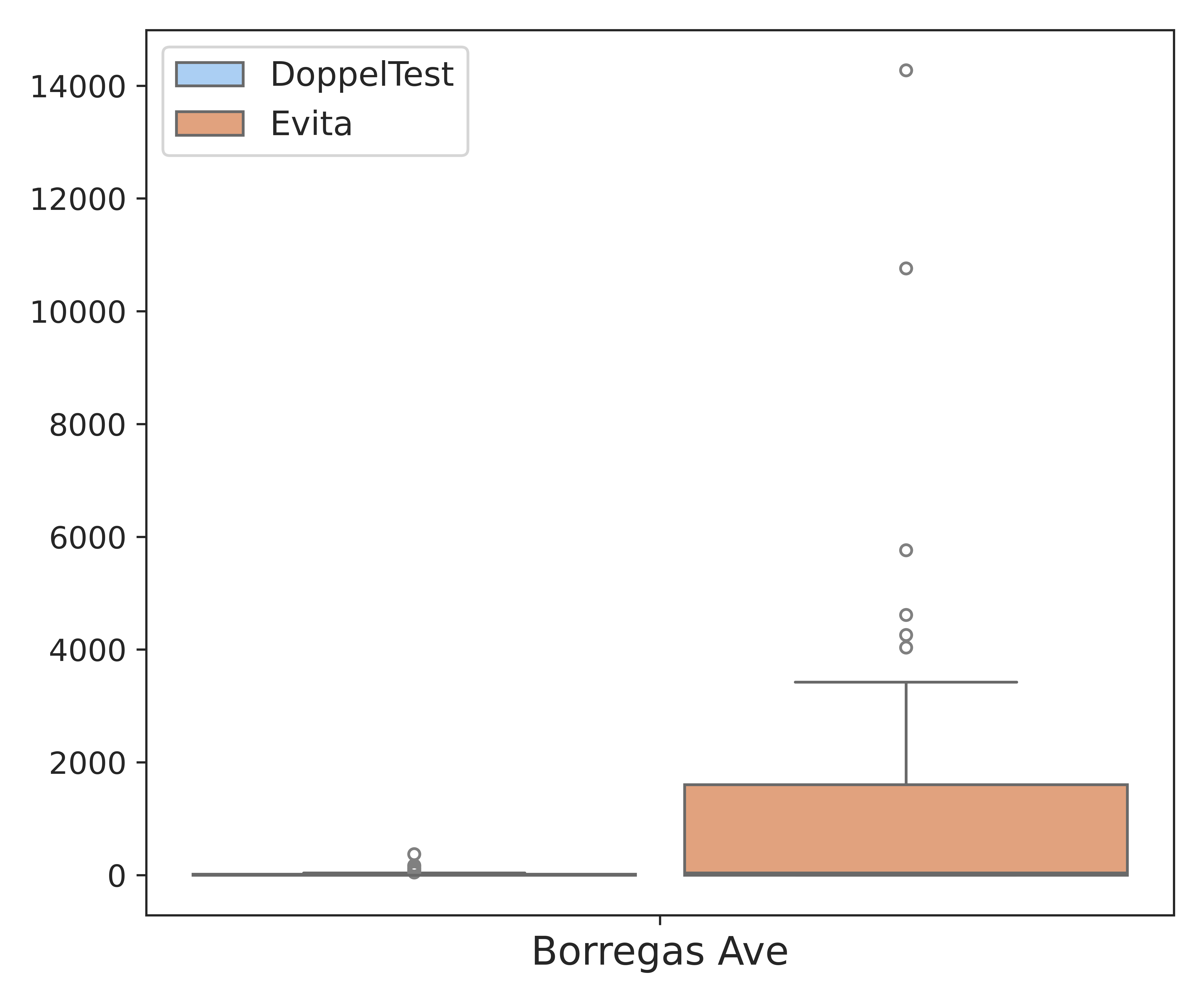}
\caption{Speeding Violations}
\label{fig:speedingViolations}
\end{subfigure}
\\
\vspace{10pt}
%
\begin{subfigure}{0.4\textwidth}
\centering
\includegraphics[width=\linewidth]{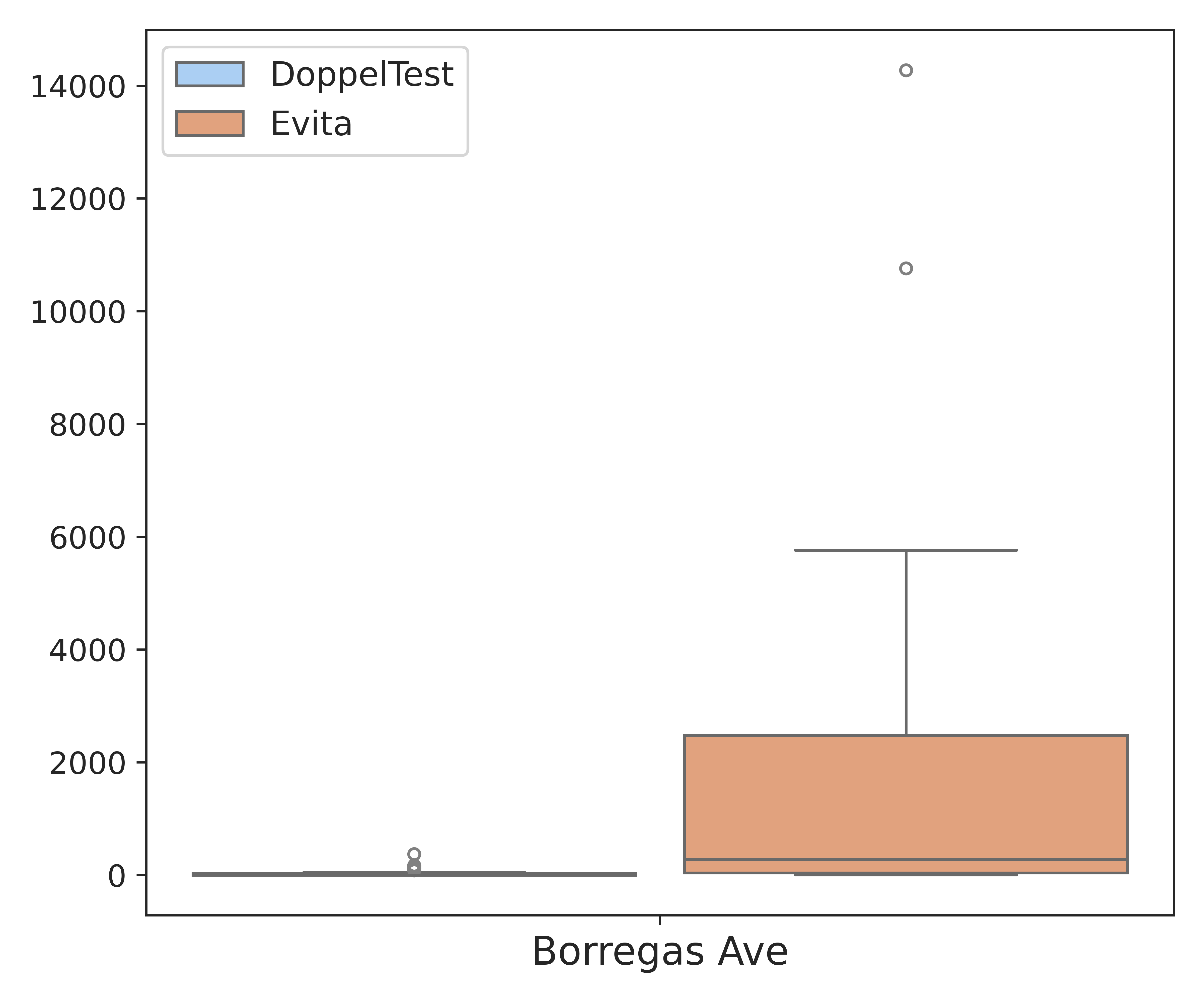}
\caption{Traffic Lights Violations}
\label{fig:trafficLightsViolations}
\end{subfigure}
\caption{RQ3 -- Number of detected violations with the \baidu experimental setting}
\label{fig:rq3numViolations}
\end{figure}

Figure~\ref{fig:rq3numViolations} shows the total number of traffic light (\totalFailTrafficLights), stop signs (\totalFailStop), and speeding (\totalSpeeding) violations detected in the scenarios generated with \approach and \doppelTest in the urban traffic environment.

Figure~\ref{fig:rq3numViolations} shows a small statistical advantage of \approach over \doppelTest for stop sign violations, and a statistically relevant advantage of \approach over \doppelTest for both speeding and traffic light violations.
\approach's advantage for both speeding and traffic light violations is particularly relevant given the poor performance of \doppelTest.

\begin{tcolorbox}[size=title, colframe=white, width=1\linewidth, colback=gray!20, breakable]
{\bf Answer to RQ3.} \approach outperforms \doppelTest across all the evaluated traffic violations, showing a good efficiency in exposing traffic rule infractions caused by \AVs.
\end{tcolorbox}

\subsection{RQ4~--~\rqFour}
\label{sec:results-rq4}

RQ4 evaluates the ability of the \approach to generate scenarios that contain only the \AVs required for triggering the interactions, thus improving testing, comprehension, and debugging.

\begin{figure}[!t]
\centering
\begin{subfigure}{0.475\textwidth}
\centering
\includegraphics[width=\linewidth]{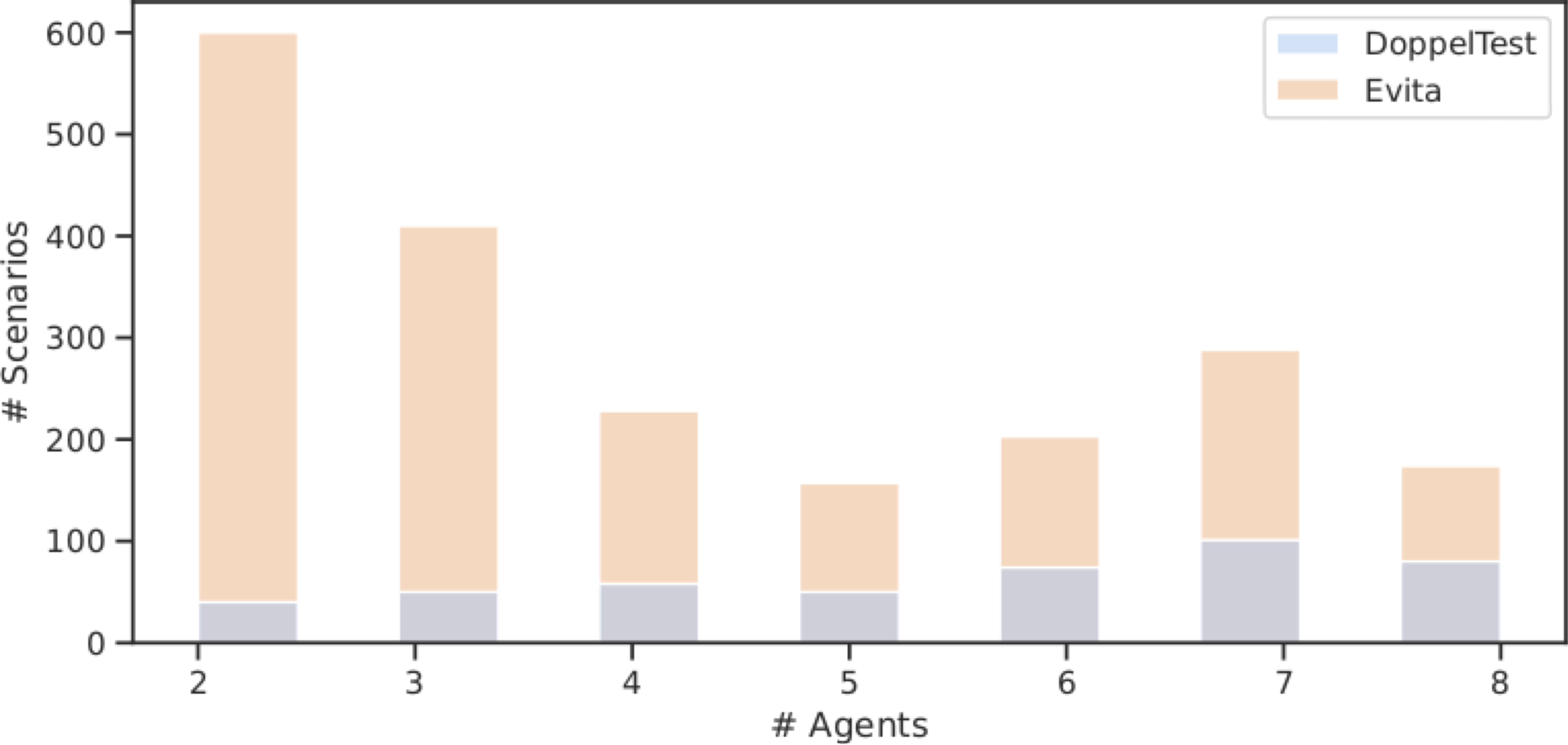}
\caption{\choID}
\label{fig:mergeScenario_numCars}
\end{subfigure}
\quad
%
\begin{subfigure}{0.475\textwidth}
\centering
\includegraphics[width=\linewidth]{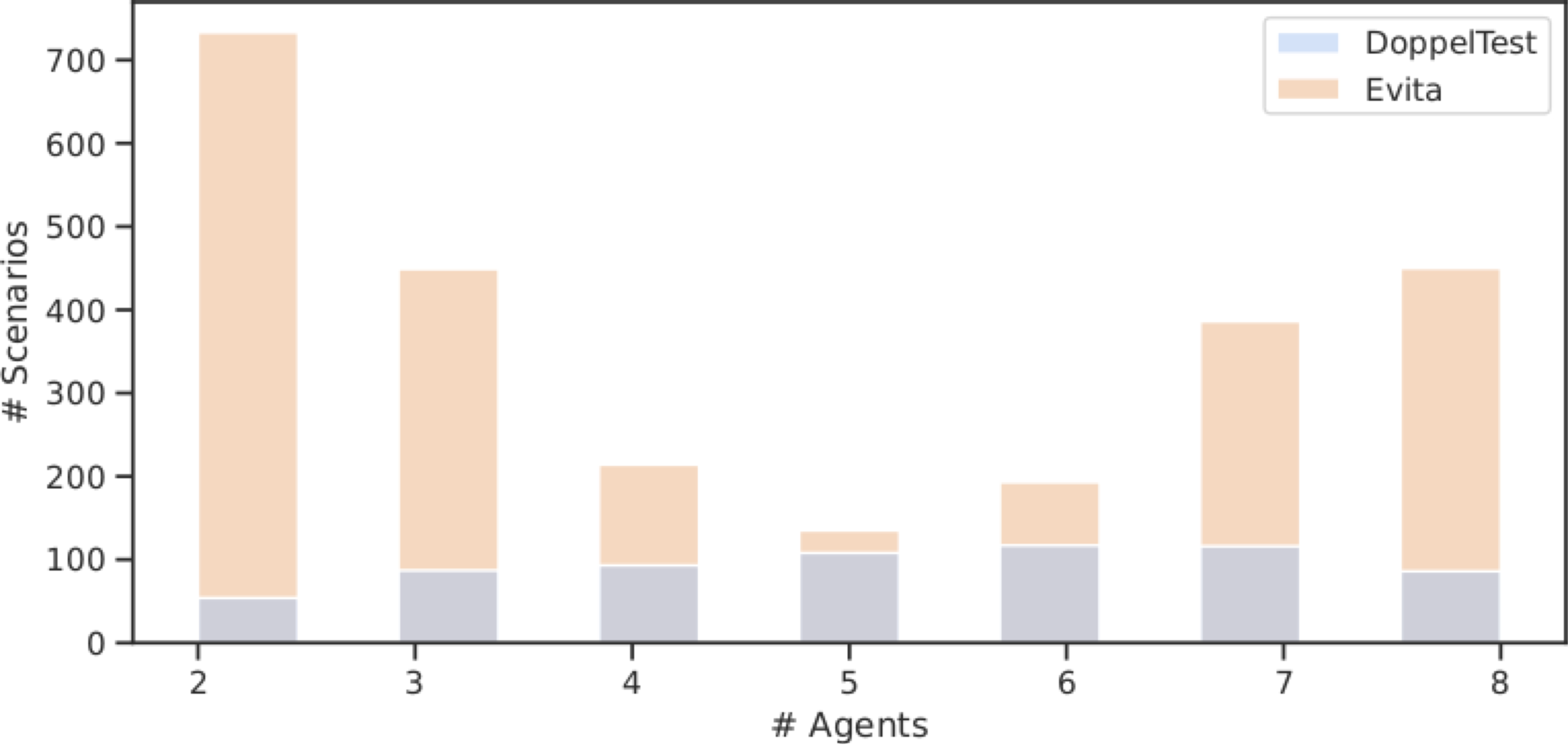}
\caption{\cologne}
\label{fig:oncomingLaneScenario_numCars}
\end{subfigure}
\\
\vspace{10pt}
\begin{subfigure}{0.475\textwidth}
\centering
\includegraphics[width=\linewidth]{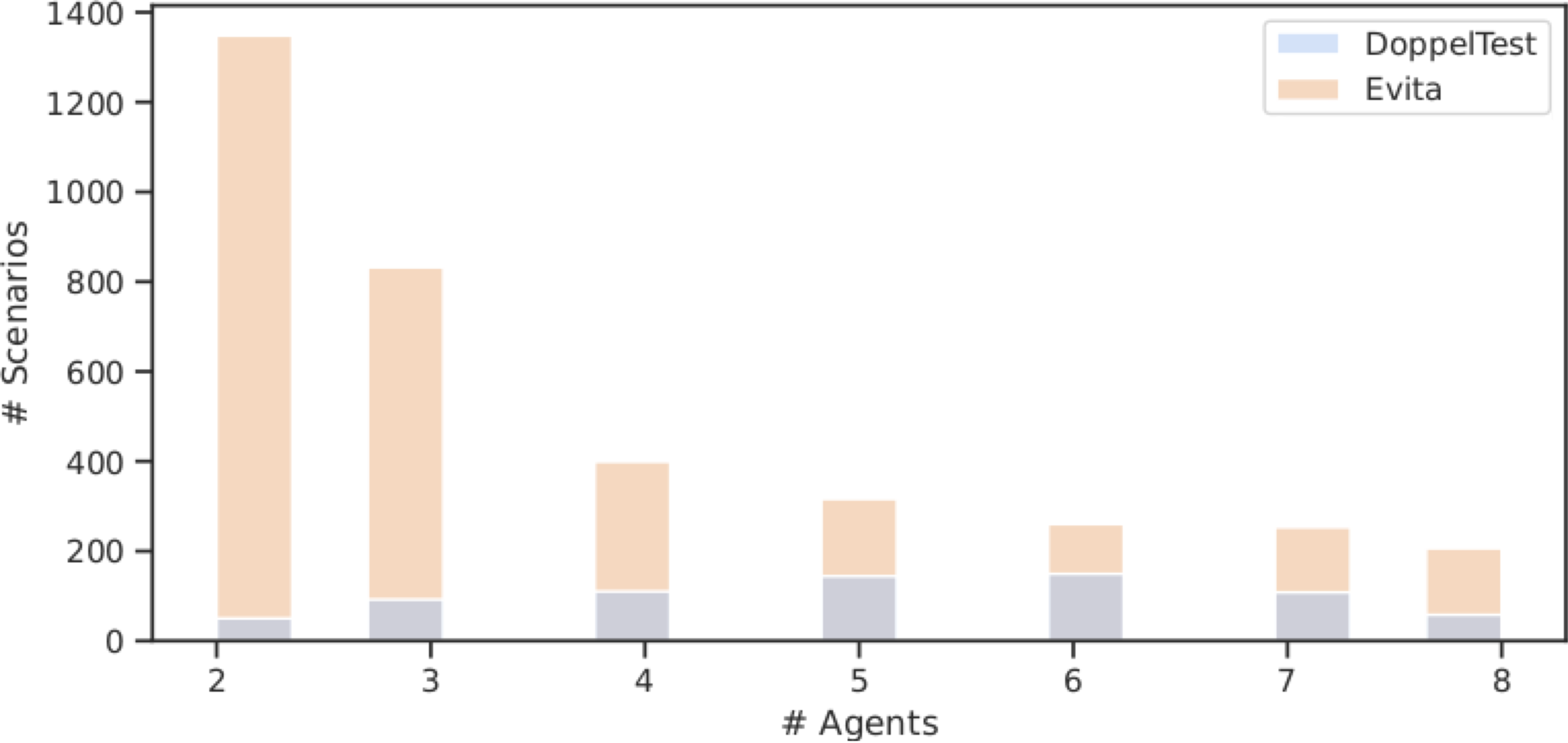}
\caption{\usaNine}
\label{fig:US101_09_numCars}
\end{subfigure}
\quad
%
\begin{subfigure}{0.475\textwidth}
\centering
\includegraphics[width=\linewidth]{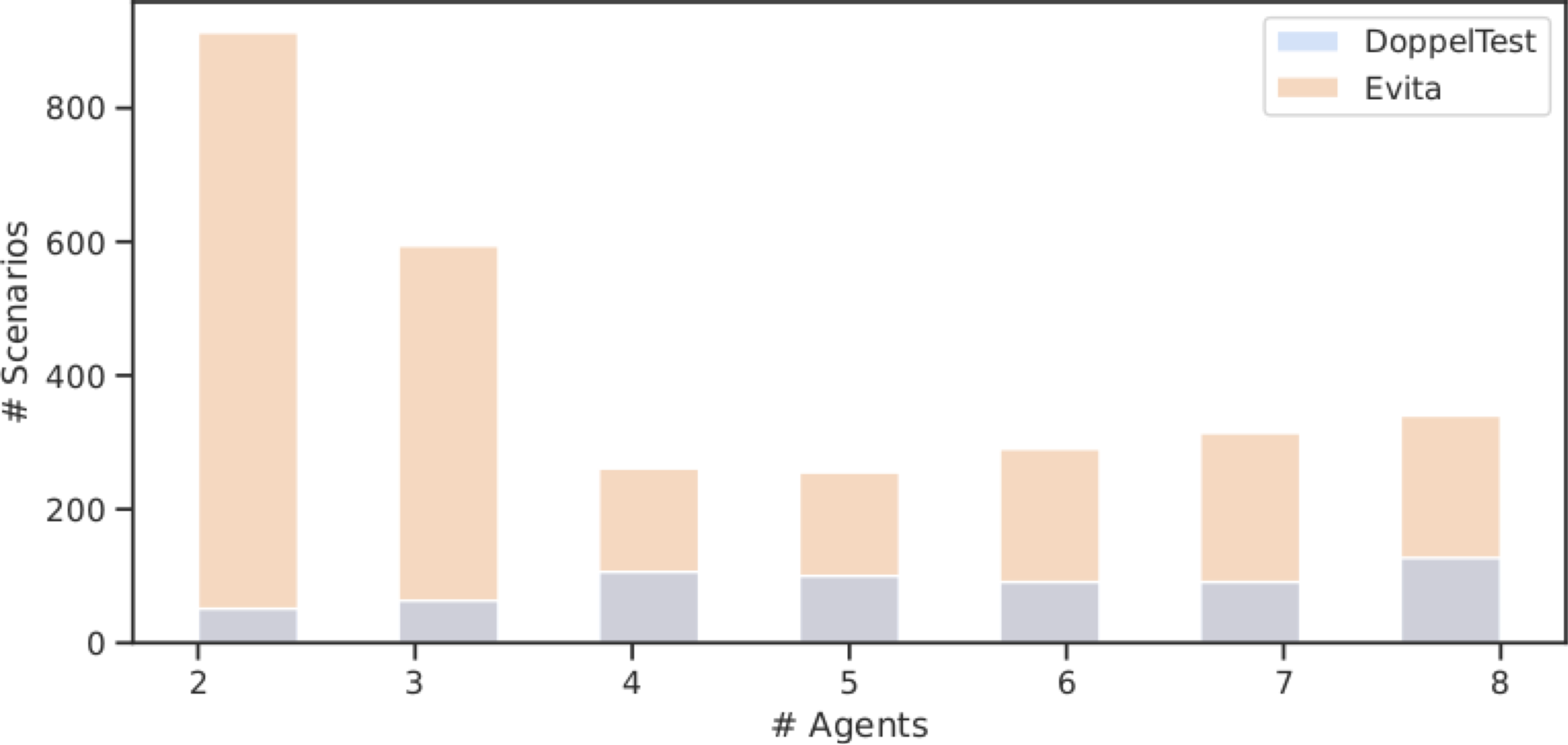}
\caption{\usaFifteen}
\label{fig:US101_15_numCars}
\end{subfigure}
\\
\vspace{10pt}
\begin{subfigure}{0.475\textwidth}
\centering
\includegraphics[width=\linewidth]{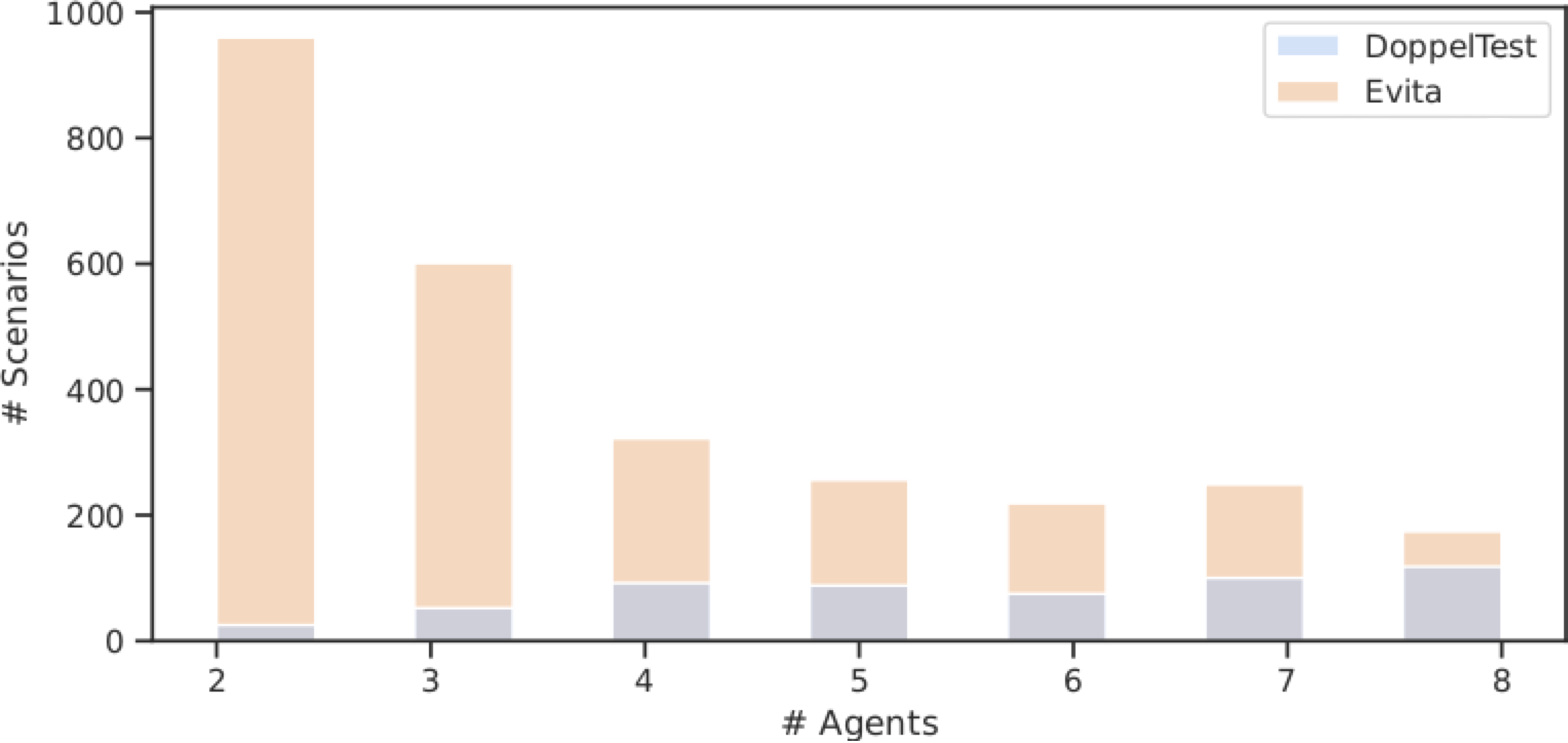}
\caption{\usaTwentythree}
\label{fig:US101_23_numCars}
\end{subfigure}
\caption{RQ4 -- Number of vehicles in the generated scenarios with the \frenetix experimental setting}
\label{fig:rq4numCars}
\end{figure}

Figure~\ref{fig:rq4numCars} reports the number of scenarios generated with \approach and \doppelTest by number \givenNumberCars of involved vehicles, $2 \leq \givenNumberCars \leq 8$, for each road network used for the \frenetix experimental setting.
In the figure, the bars representing the number of scenarios for the two generators are overlaid.
Figure~\ref{fig:rq4numCars} shows that \approach generates simpler scenarios than \doppelTest. 
Both approaches generate scenarios with any number  \givenNumberCars of vehicles, $2 \leq \givenNumberCars \leq 8$.
However, \approach always generates a majority of scenarios with small number \givenNumberCars of vehicles, $2 \leq \givenNumberCars \leq 4$, whereas \doppelTest generates a majority of scenarios with high number \givenNumberCars of vehicles, $5 \leq \givenNumberCars \leq 8$. 

Both \approach and \doppelTest generate a significantly high number of scenarios with many vehicles ($7$ and $8$) for the road networks \choID and \cologne.
We argue that complex road networks require multiple vehicles to generate a wide diversity of interactions.

\medskip

\begin{figure}
\centering
\includegraphics[width = 0.575\linewidth]{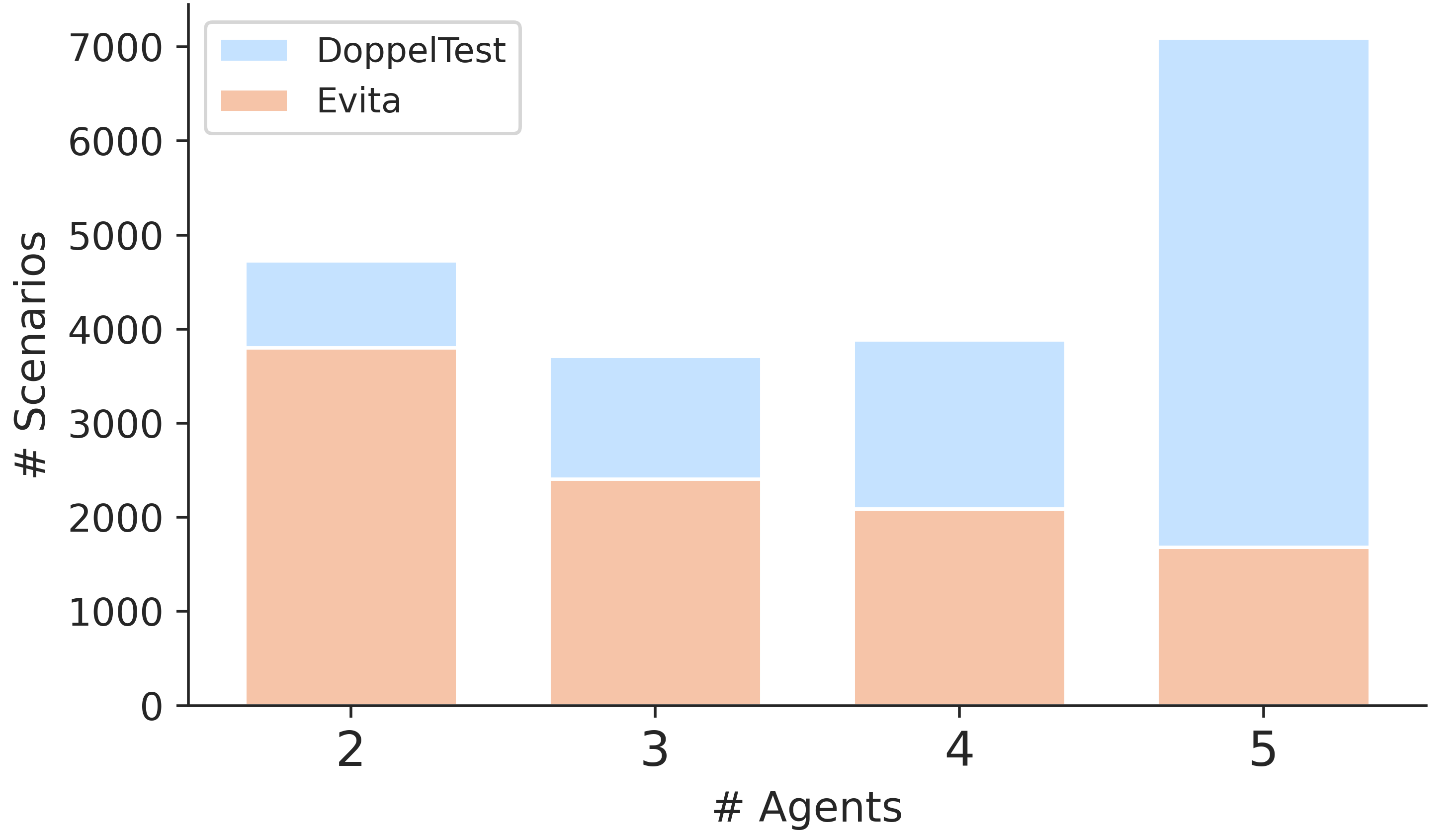}
\caption{RQ4 -- Number of vehicles in the generated scenarios with the \baidu experimental setting}
\label{fig:rq4numCarsBorregas}
\end{figure}

Figure~\ref{fig:rq4numCarsBorregas} shows the distribution of scenarios based on the number \givenNumberCars of vehicles, $2 \leq \givenNumberCars \leq 5$ for the \baidu urban road network.
In the figure, the bars representing the approaches are overlaid. 
Figure~\ref{fig:rq4numCarsBorregas} confirms the results in the \frenetix highway and suburban traffic (Figure~\ref{fig:rq4numCars}): \approach generates a higher percentage of scenarios with few vehicles than \doppelTest.

These results confirm the results of RQ1: \approach systematically triggers critical interactions, collisions, and traffic rule violations while keeping the test cases simple and lightweight, while \doppelTest relies on complex scenarios with several vehicles to uncover unique critical situations.

\begin{tcolorbox}[size = title, colframe = white, width = \linewidth, colback = gray!20, breakable]
{\bf Answer to RQ4.} \approach generates scenarios with fewer vehicles than \doppelTest.
It generates a high number of scenarios with many vehicles when required to obtain a wide range of \AVs interactions in complex road networks.
\end{tcolorbox}

\subsection{RQ5~--~\rqFive}\label{sec:results-rq5}
\begin{figure*}[t]
\centering
\begin{subfigure}{0.65\linewidth}
\centering
\includegraphics[width=\linewidth]{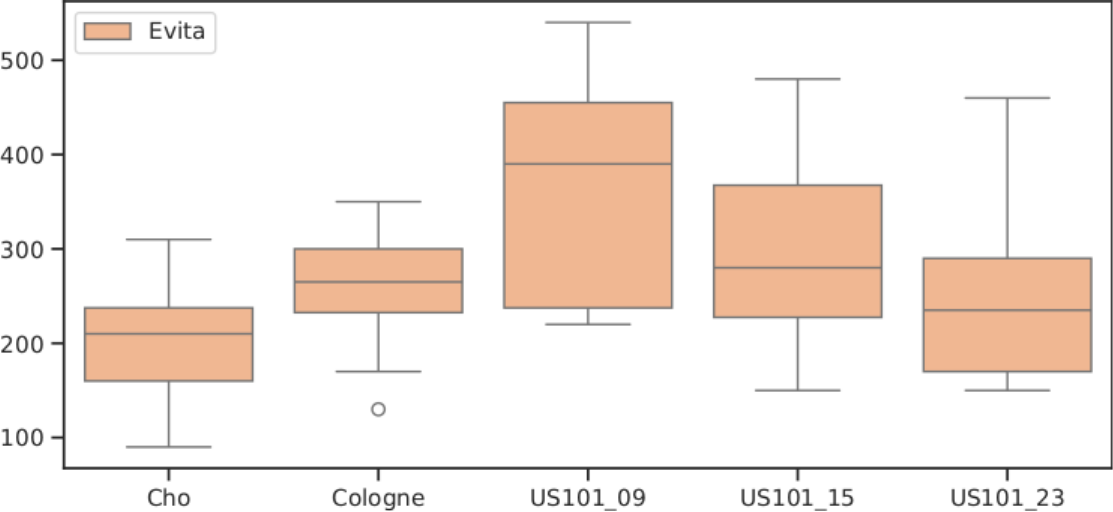}
\caption{Total generated scenarios}
\label{fig:rq4genStatsAll}
\end{subfigure}
\\
\vspace{10pt}
\begin{subfigure}{0.64\linewidth}
\centering
\includegraphics[width=\linewidth]{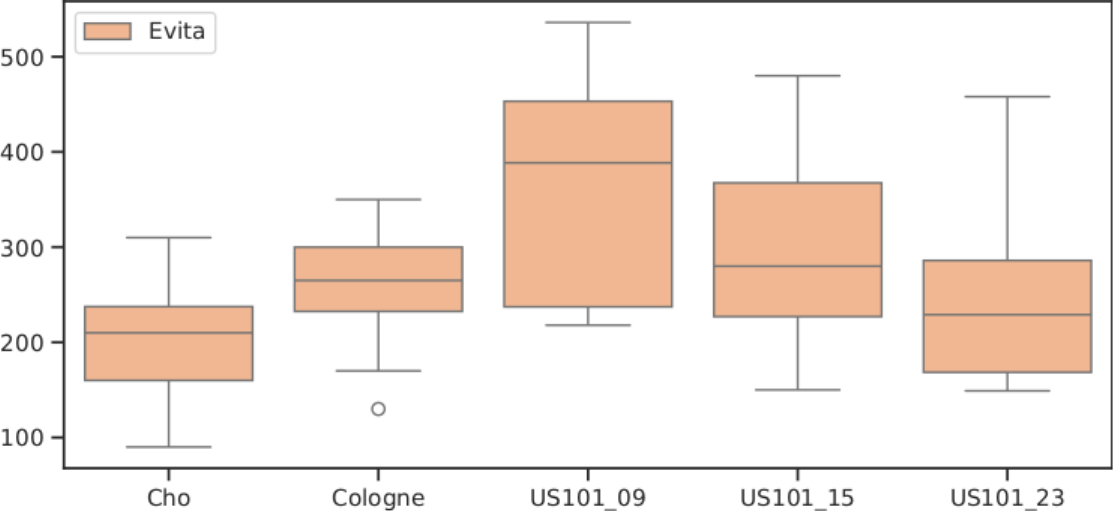}
\caption{Valid scenarios}
\label{fig:rq4genStatsValid}
\end{subfigure}
\caption{RQ5 -- Generation statistics with the \frenetix experimental setting}
\label{fig:rq5genStats}
\end{figure*}

RQ5 assesses the effectiveness of \approach as a test generator, in terms of the percentage of valid scenarios that \approach generates.
Figure~\ref{fig:rq5genStats} reports the number of total and valid scenarios that \approach generates for each \frenetix road network.
The almost complete overlapping of the two figures shows that \approach generates a negligible number of invalid scenarios. 
\approach generates between $200$ and $400$ scenarios on average for each road network with a budget of $3$ CPU-hours.
It generates a scenario in $27$ to $34$ seconds, depending on the complexity of the network. 

\begin{figure*}[t]
\centering
\begin{subfigure}{0.45\linewidth}
\centering
\includegraphics[width = \linewidth]{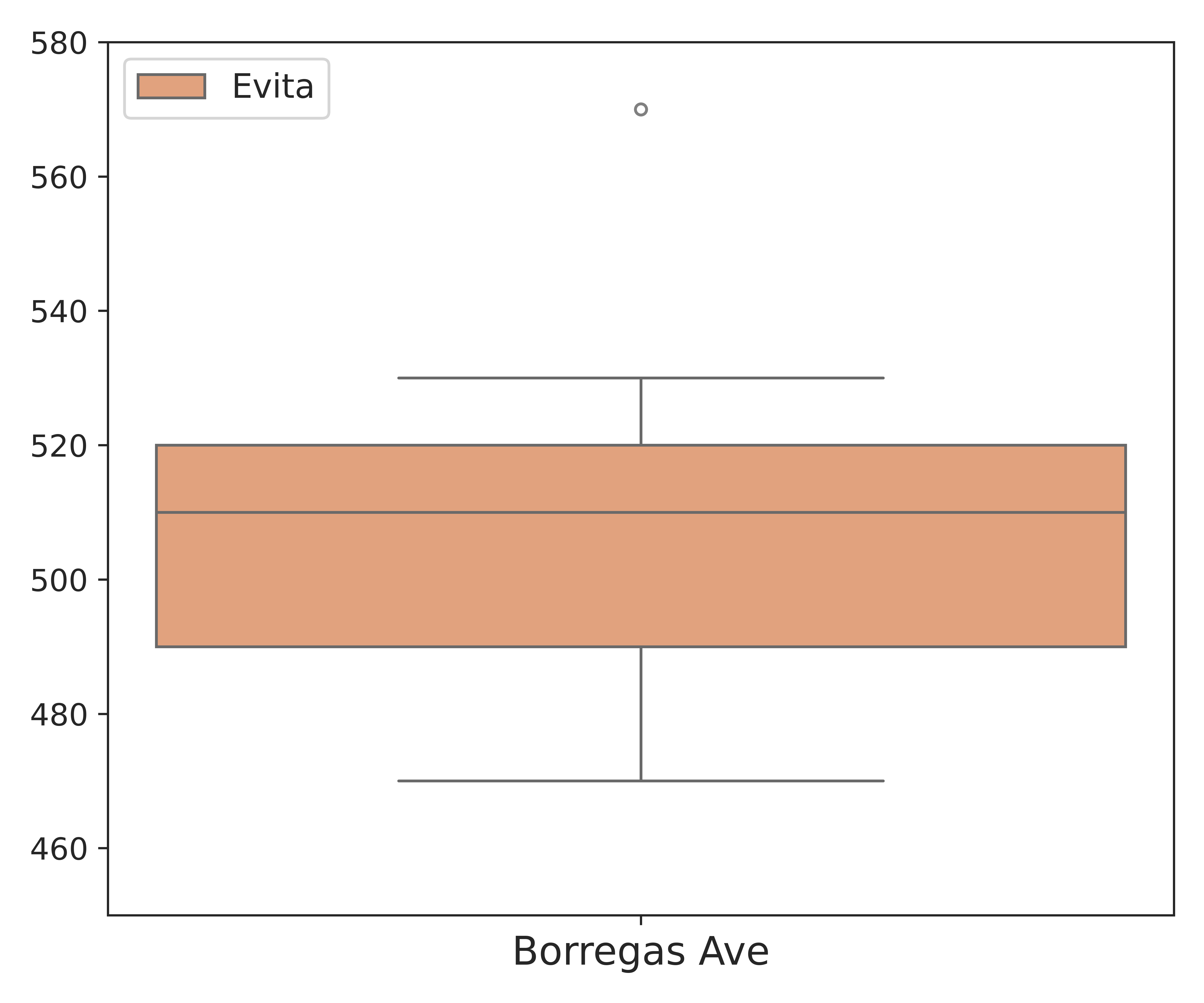}
\caption{All generated scenarios}
\label{fig:rq4genStatsAllBorregas}
\end{subfigure}
\begin{subfigure}{0.45\linewidth}
\centering
\includegraphics[width = \linewidth]{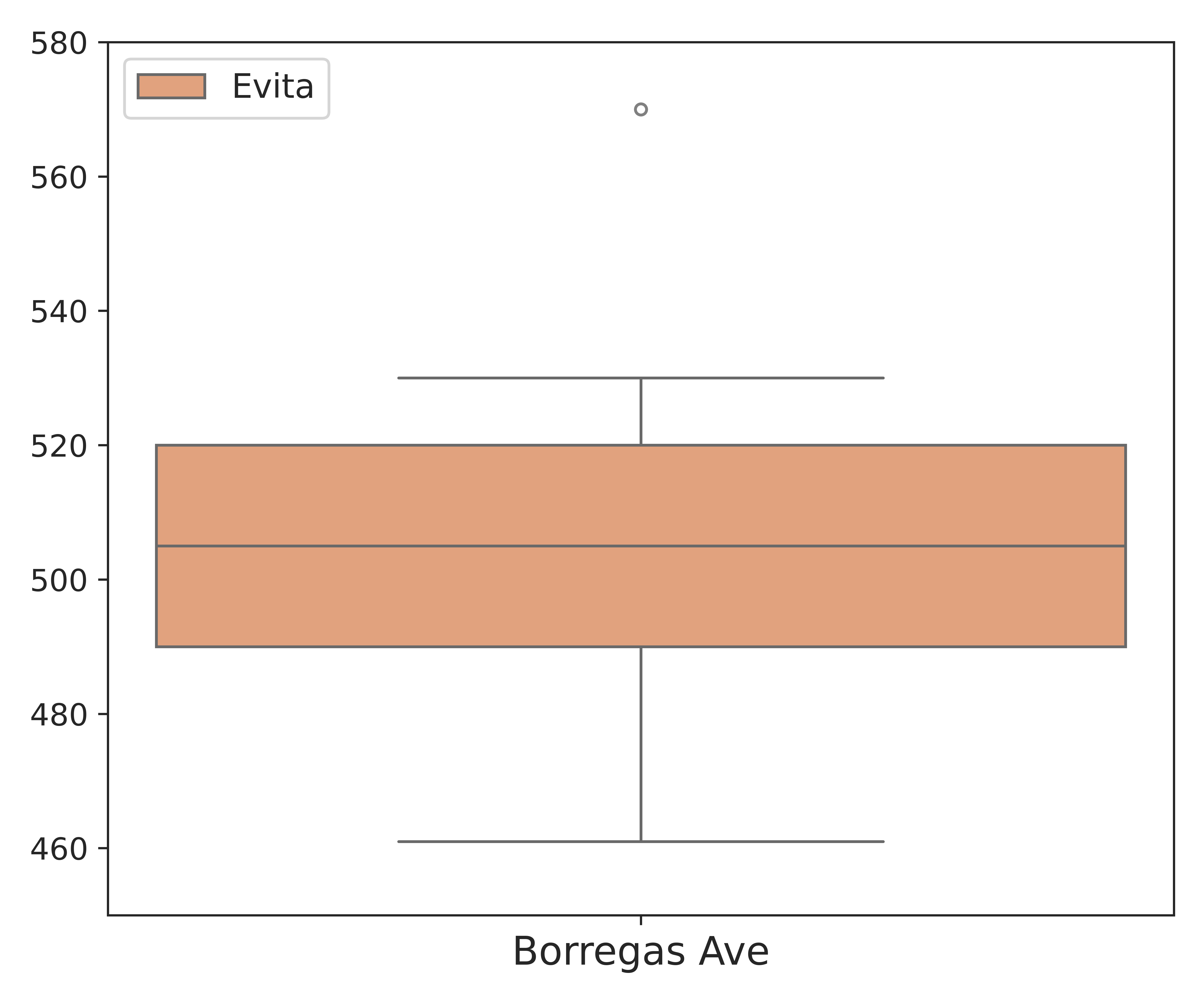}
\caption{Valid scenarios}
\label{fig:rq4genStatsValidBorregas}
\end{subfigure}
\caption{RQ5 -- Generation statistics with the \baidu experimental setting}
\label{fig:rq5genStatsBorregas}
\end{figure*}

\smallskip


Figure~\ref{fig:rq5genStatsBorregas} shows the number of total and valid scenarios that \approach generates for the \baidu experimental setting.
The near overlapping of the two diagrams confirms the results: \approach generates a negligible number of invalid scenarios ($29$ out of $490$ in the worst case).    
\approach generates $470$ to $570$ scenarios on average for the \borregas road network with a budget of $12$ CPU-hours. It generates a scenario in $76$ to $92$ seconds.
The generation time for the scenarios of the \baidu network is higher than the \frenetix network, due to the parsing operations required to convert between \CommonRoad and \apollo compatible scenarios.

\begin{tcolorbox}[size = title, colframe = white, width = \linewidth, colback = gray!20, breakable]
{\bf Answer to RQ5.} \approach is effective as it generates scenarios quickly, and only a negligible number of them are invalid. 
\end{tcolorbox}

\section{Threats to Validity}
\label{sec:threatsToValidity}
We present an overview of the main construction, internal, and external validity threats, and discuss how we mitigated them.

\subsection{Construction Validity}
The main threats to the construction validity stem from how we check the interactions and the different settings of \approach and \doppelTest in the comparative evaluation.

\paragraph{Interaction checks}
In the experimental setup for testing \frenetix, \approach identifies interactions as changes in consecutively planned trajectories coupled with the presence of specific scenario elements (vehicles, traffic lights, stop signs) within a safety distance (see conditions \firstCondInteration and \secondCondInteration in Section~\ref{sec:interactions}).
However, a change in a planned trajectory does not necessarily originate from an interaction with such scenario elements; thus, we may suffer from some false positives. 
We opted for efficiency over precision and chose a metric that can be quickly computed to guide scenario generation, since it is on the critical path of \approach. 
The search algorithm at the core of our approach is robust and works well with approximate metrics to drive generation towards exposing multiple, diverse interactions.  Our sample inspection of the results indicates a negligible impact of false positives on the presented results.

\paragraph{Settings}
\doppelTest generates scenarios with an initial velocity of the \AVs of $0$~km/h, while \approach generates scenarios with any initial non-negative velocity for the \AVs.
This difference may bias the comparison of the two approaches.
We executed all experiments for \approach with an initial velocity of $0$~km/h for all \AVs to check for potential bias.
The experiments confirm the results reported in Section~\ref{sec:results}. 

\subsection{Internal Validity}
The main threats to the internal validity come from the randomness of the search algorithms and the quality of the prototype.

\paragraph{Randomness of the search algorithms}
The randomness of search algorithms could lead to misleading results.
To mitigate this threat, we followed established guidelines~\cite{ArcuriICSE11}: We repeated the experiments multiple times (10 repetitions) and analyzed the results using statistical tests that consider statistical significance and effect size.

\paragraph{Quality of the framework}
Possible bugs in the implementation of \approach and \doppelTest extensions can undermine the validity of the experiments, for instance, by failing to detect a collision between two vehicles or by recording metrics imprecisely.
We thoroughly tested the framework implementation, and we provide access to a replication package~\cite{replication-package-evita} to further validate the framework and the experiments.

\subsection{External Validity}
The main threat to external validity is the limited scope of the experiment, which limits the generalizability of the results.
We mitigate this risk by considering two simulators and test subjects (\frenetix and \baidu) and six road networks (\choID, \cologne, \usaNine, \usaFifteen, \usaTwentythree, \borregas).
We modeled the road networks after real road segments, with representative road features, and a wide range of driving scenarios.

\section{Related Work}
\label{sec:related}
The many recent surveys on \AVs testing~\cite{Wang:AVTestingScenarioGeneration:2024, Jinkang:ScenarioGenerationAV:IV:2024, AVTesting-Survey,DBLP:conf/sigsoft/LouDZZ022} agree that most current approaches focus on testing individual \AVs, and exploit search-based algorithms, fuzzing, or artificial intelligence to explore the testing scenario space.
Ben~Abdessalem et al.~\cite{Abdessalem2016ASE,Abdessalem2018TVC} combine surrogate and machine-learning models with search-based algorithms to test the safety of some systems, such as the pedestrian detection and emergency braking systems.
Gambi et al.~\cite{DBLP:conf/issta/GambiMF19} propose search-based procedural content generation to generate road structures for testing the lane-keeping systems.
Luo et al.~\cite{reqsViolADSsASE21} propose a search-based approach to find scenarios that expose different combinations of traffic violations.

Other approaches rely on fuzzing. Li et al.~\cite{LiISSRE2020} propose AV-FUZZER, a fuzzing approach to find collisions. Cheng et al.~\cite{BehAVExplor} propose a behavior-guided fuzzing technique to find different behaviors of the \egovehicle that can potentially lead to different types of violations.
The most recent test generator approaches exploit Reinforcement Learning and Generative AI. Lu et al.~\cite{LuTSE2023} use Reinforcement Learning to find environments in which the vehicle collides.
Tang et al.~\cite{TangASE24} use LLMs to transform collision reports into critical scenarios.
In this paper, we propose search-based algorithms to generate test cases for testing non-cooperative interacting autonomous vehicles.

Only few approaches target the testing of interacting autonomous components.
Ben Abdessalem et al.~\cite{Abdessalem2018TAC} generate tests that expose feature interactions, which they define as unforeseen interactions among the autonomous modules comprising the \av, for instance, cruise control and emergency braking. \approach targets the interactions among multiple \AVs.

Gambi et al.~\cite{testInAVsIV2025} and Huai et al.~\cite{doppleganger} generate critical scenarios involving multiple \AVs.
Gambi et al.'s TIAV~\cite{testInAVsIV2025} is a single-objective genetic algorithm that identifies critical highway and suburban scenarios by maximizing danger~\cite{avoidCollICST2020}.
Huai et al.'s \doppelTest~\cite{doppleganger} is a multi-objective genetic algorithm that identifies scenarios in which the \AVs violate traffic rules.
\approach, TIAV and \doppelTest share the same basic objective: testing multiple \AVs to identify collisions that testing a single \AVs misses. 
\approach improves the encouraging results of TIAV and \doppelTest by emphasizing the various kinds of \AVs' interactions, and exploring the space of \AVs interactions to expose a wide variety of critical situations and traffic violations. \approach generates highway, suburban, and urban scenarios and does not make assumptions on the ability to control dynamic obstacles such as pedestrians.

Interactions are closely related to the more general concept of causality~\cite{pearl2009causality}, a vast research field that aims to understand how an event (the cause) brings about another event (the effect).
While causality has recently been used in the context of testing AVs~\cite{GiamatteiTOSEM2024}, its definition remains quite abstract and is often impractical and expensive to compute. 
To the best of our knowledge, our definition is the first lightweight definition of interactions that is both effective in describing interactions between vehicles in autonomous driving and efficient to compute.

\section{Conclusions and Future Work}
\label{sec:conclusions}
In this paper, we propose \approach, an approach for testing the safety of \AVs interactions, identifying safety-critical scenarios, and checking \AVs under a large variety of types of interactions. 
While state-of-the-art approaches~\cite{Huang2016,AVTesting-Survey} test single \AVs reacting to pre-programmed NPCs, and thus miss critical issues that emerge from the interactions of independent \AVs, \approach tests critical scenarios that involve multiple interacting \AVs with an original multi-objective search-based approach.

The results of our experiments that we report in the paper confirm that generating scenarios involving multiple interacting \AVs can effectively expose a wide range of safety failures in \AVs.
Our experiments with two popular simulators and different road networks confirm the generality of  \approach. 

\approach is a relevant step towards ensuring the quality of autonomous driving at scale.
It provides a solid foundation and a ready-to-use framework for further investigating \AVs interactions, thereby opening up interesting research directions.

The results presented in this paper contribute to the research on \AVs by
\begin{inparaenum}[(i)]
\item providing a precise and accurate definition of \AVs' interactions that enables \approach to prune irrelevant interactions during test generation, thus boosting the efficiency and effectiveness of test generation,
\item extending the scope of testing \AVs towards testing the interactions among \AVs from different manufacturers, and 
\item discussing the results obtained by experimenting in different traffic conditions and with diverse static and dynamic traffic obstacles.
\end{inparaenum}


\Urlmuskip=0mu plus 1mu\relax
\bibliographystyle{ACM-Reference-Format}
\bibliography{biblio}

\end{document}